\documentclass[aps,pra,twocolumn,showpacs,amsmath,amsmath,amssymb,superscriptaddress,longbibliography]{revtex4-1}

\usepackage{graphicx}
\usepackage{dcolumn}
\usepackage{bm}
\usepackage{layouts}
\usepackage{dsfont}
\usepackage{amssymb}
\usepackage{bbold}
\usepackage{float}

\newcommand{\mi}{\mathrm{i}}

\newcommand{\bra}[1]{\ensuremath{\langle{#1}|}}
\newcommand{\ket}[1]{\ensuremath{|{#1}\rangle}}

\makeatletter
\newcommand\thefontsize[1]{{#1 The current font size is: \f@size pt\par}}
\makeatother

\begin{document}

\title{Subradiance-protected excitation transport}

\author{Jemma A. Needham}
\author{Igor Lesanovsky}	
\author{Beatriz Olmos}
\email{beatriz.olmos-sanchez@nottingham.ac.uk}
\affiliation{School of Physics and Astronomy, The University of Nottingham, Nottingham, NG7 2RD, United Kingdom}
\affiliation{Centre for the Mathematics and Theoretical Physics of Quantum Non-equilibrium Systems, The University of Nottingham, Nottingham, NG7 2RD, United Kingdom}

\date{\today}
	
\begin{abstract}
We explore excitation transport within a one-dimensional chain of atoms where the atomic transition dipoles are coupled to the free radiation field. When the atoms are separated by distances smaller or comparable to the wavelength of the transition, the exchange of virtual photons leads to the transport of the excitation through the lattice. Even though this is a strongly dissipative system, we find that the transport is subradiant, that is, the excitation lifetime is orders of magnitude longer than the one of an individual atom. In particular, we show that a subspace of the spectrum is formed by subradiant states with a linear dispersion relation, which allows for the dispersionless transport of wave packets over long distances with virtually zero decay rate. Moreover, the group velocity and direction of the transport can be controlled via an external uniform magnetic field while preserving its subradiant character. The simplicity and versatility of this system, together with the robustness of subradiance against disorder, makes it relevant for a range of applications such as lossless energy transport and long-time light storage.
\end{abstract}
	
\maketitle

\section{Introduction}

An ensemble of emitters couples collectively to a common electromagnetic bath, as was already investigated theoretically in the seminal papers of Dicke, Lehmberg and Agarwal in the 1950s and 70s \cite{Dicke1954,Lehmberg1970,Agarwal1971}. Here, the exchange of virtual photons results in induced dipole-dipole interactions \cite{Olmos2013,Sutherland2016,Bettles2016} and collective Lamb and Lorentz-Lorenz shifts \cite{Friedberg1973,Scully2009,Keaveney2012,Pellegrino2014,Jenkins2016,Bromley2016,Peyrot2018}. Moreover, the emission of photons into the bath takes place at a rate much faster or slower (so-called super- and subradiance, respectively) than the single atom decay rate \cite{Bienaime2012,Ott2013,Oliveira2014,Guerin2016,Araujo2016,Roof2016,Araujo2017,Cottier2018}. This cooperative behavior is featured both in dense ensembles, where the interparticle separations are comparable to the wavelength of the scattered light, and in dilute ones with a very large number of emitters. Super- and subradiance have been observed experimentally not only in atomic gases, but also in QED circuits \cite{Filipp2011,Lalumiere2013}, metamaterial arrays \cite{Jenkins2017}, and quantum dots \cite{Lodahl2004,Tighineanu2016,Kim2018}. This collective atom-light coupling has found a variety of applications such as storage of light via the preparation of subradiant states through phase-imprinting protocols \cite{Scully2006,Plankensteiner2015,Scully2015,Jen2016,Facchinetti2016,Asenjo-Garcia2017,Jen2017,Jen2018,Jen2018-1,Moreno2019,Guimond2019}, topologically protected transport of excitations \cite{Bettles2017,Perczel2017}, or efficient long-range energy transport \cite{Willingham2011,Giusteri2015,Leggio2015,Doyeux2017}.

\begin{figure}[t!]
\begin{center}		
\includegraphics[width = \columnwidth]{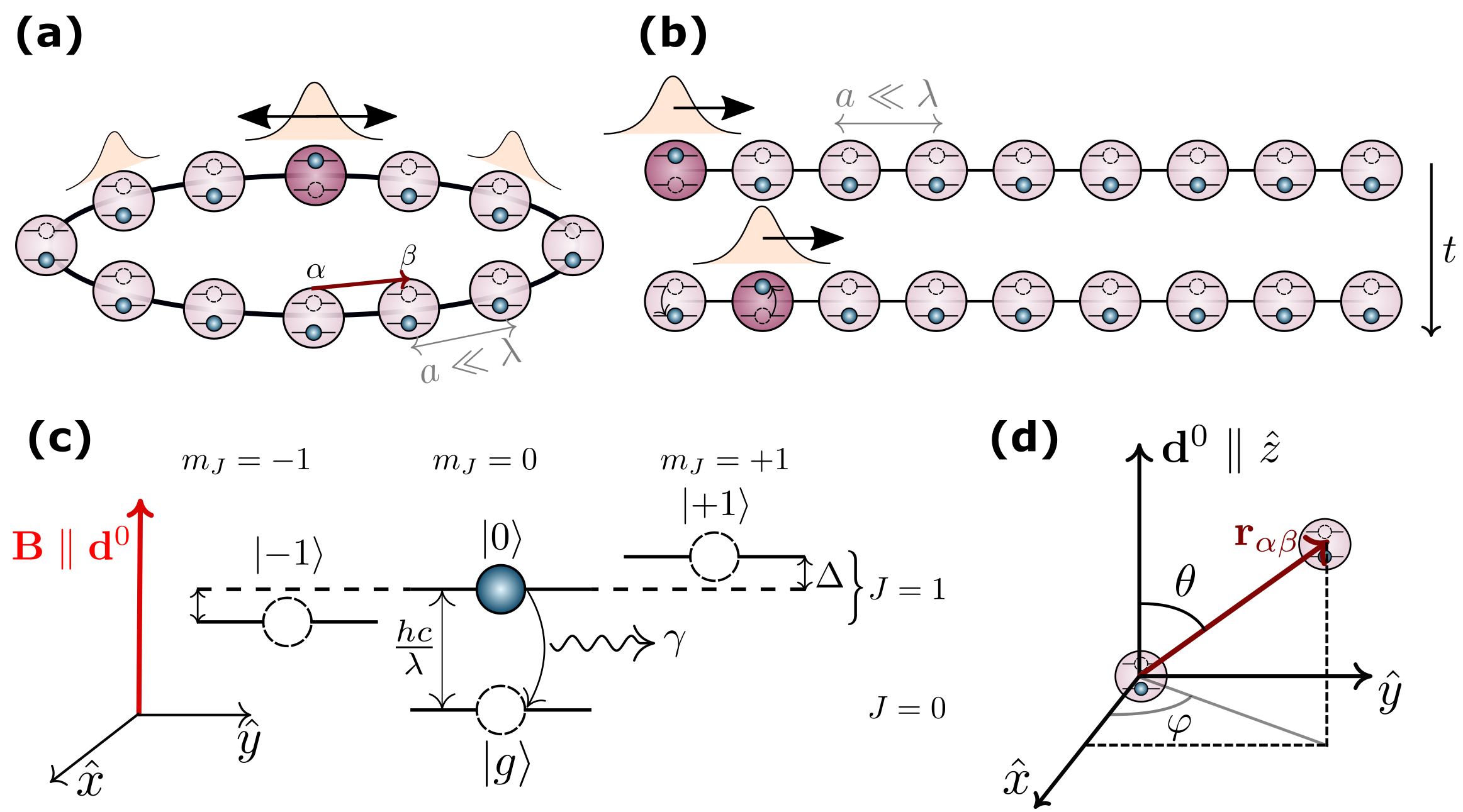}
\end{center}	
\caption{\textbf{The system.} We consider a one-dimensional chain of atoms with nearest neighbor separation $a$ in \textbf{(a)} a ring and \textbf{(b)} a linear chain with open boundary conditions. The wave packet that contains the excitation is transported via dipole-dipole interactions induced by the collective coupling to the radiation field at zero temperature. \textbf{(c):} We consider the following internal atomic levels in each atom: a ground state $\left|g\right>$ and three degenerate excited states $\left|-1\right>,\left|0\right>,\left|+1\right>$. The degeneracy is lifted by the shift of $\left|\pm1\right>$ by $\pm\Delta=\pm\mu_Bg\left|\mathbf{B}\right|$ when an external uniform magnetic field $\mathbf{B}$ is aligned with the dipole moment $\mathbf{d}^0$ of the $\left|g\right>\to\left|0\right>$ transition (quantization axis). \textbf{(d):} The polar angle $\theta$ of $\mathbf{r}_{\alpha\beta}$ (separation between the $\alpha$-th and $\beta$-th atoms) controls the strength of the interactions and the collective character of the dissipation.}
\label{fig:Fig1}
\end{figure}

In this paper, we show that it is possible to realize subradiance-protected transport of a wave packet through a dense atomic chain with lifetimes many orders of magnitude longer than the one of an individual atom. This is achieved by maximizing the overlap of the wave packet with a subradiant manifold of states that possess a linear dispersion relation. Further control over the transport can be attained by effectively changing the orientation of the transition dipoles via an external uniform magnetic field. In particular, we show that the group velocity of the wave packet can be brought close to zero while preserving its long lifetime. Finally, we analyse the effect of disorder, which arises from the width of the external wavefunction of the atoms in each lattice trap and is inevitable in a realistic experimental scenario. Even though this can lead to the suppression of the transport of the wave packet due to localization \cite{Deng2018,Botzung2018}, we find that the subradiant character of the dynamics is robust against the presence of disorder \cite{Akkermans2013,Biella2013}.

\section{Interaction between the atoms and the radiation field}

\subsection{Master Equation}

We consider an ensemble of $N$ atoms at positions $\mathbf{r}_\alpha$ with $\alpha = 1,\dots,N$, each one tightly trapped in the sites of a one-dimensional lattice with lattice constant $a$ [see Figs. \ref{fig:Fig1}(a) and (b)]. The internal degrees of freedom of each atom are considered as a generic $J=0\to 1$ transition, with a single ground state $\left|g\right>$ and three degenerate excited states $\left|-1\right>$, $\left|0\right>$ and $\left|+1\right>$. The energy difference between the ground and excited states is denoted by $\hbar\omega=hc/\lambda$, where $\lambda$ is the wavelength of the transition. We choose the transition dipole moment $\mathbf{d}^0$ of the $\left|g\right>\to\left|0\right>$ transition to be aligned with the quantization axis ($z$-axis) [Fig. \ref{fig:Fig1}(c)].

The atoms are in contact with the radiation field, which we model as a thermal bath at zero temperature, whose degrees of freedom are traced out. Within the Born-Markov and secular approximations \cite{Dicke1954,Lehmberg1970,Agarwal1971}, the master equation for the dynamics of the internal degrees of freedom encoded in the reduced density matrix $\rho$ yields
\begin{equation}
    \dot{\rho}(t) = -\frac{\mi}{\hbar}[H,\rho] + \mathcal{D}(\rho).
    \label{equ:mastereq}
\end{equation}
The Hamiltonian $H$ describes the coherent long-ranged exchange of virtual photons among the atoms and is given by
\begin{equation}
	H = \hbar\sum_{\alpha \neq \beta} \mathbf{d}_{\alpha}'\cdot\bar{V}_{\alpha\beta}\mathbf{d}_{\beta},
	\label{equ:hamiltonian}
\end{equation}
where \(
	\mathbf{d}_{\alpha} = \begin{pmatrix}
	d_{\alpha}^{+1} &
	d_{\alpha}^{0} &
	d_{\alpha}^{-1} \\
	\end{pmatrix}^T\) and \(\mathbf{d}_{\alpha}^{'} = \begin{pmatrix}
	d_{\alpha}^{+1\dagger} &
	d_{\alpha}^{0\dagger} &
	d_{\alpha}^{-1\dagger} \\
	\end{pmatrix}^T\) with the atomic lowering and raising operators defined as $d_{\alpha}^{m} = \ket{g}_\alpha\bra{m}\) and \({d^{m\dag}_{\alpha}} = \ket{m}_\alpha\bra{g}$, respectively, for $m = -1,0,+1$ and $\alpha = 1,\dots,N$. The coherent exchange rate between two atoms $\alpha$ and $\beta$ is represented by the coefficient matrix
	\begin{equation}
		\bar{V}_{\alpha\beta} = \frac{3\gamma}{8}\begin{pmatrix}
V_{11}&-V_{10}^*&V_{+-}^*\\ 
-V_{10} &V_{00}&V_{10}^*\\ 
V_{+-}&V_{10}&V_{11}\\
\end{pmatrix},\label{eq:V}
	\end{equation}
with 
\begin{eqnarray*}
&&V_{11}=\left(2-3\sin^2{\theta}\right)\!\!\left(\frac{\sin{\kappa}}{\kappa^2} + \frac{\cos{\kappa}}{\kappa^3}\right)\!\!-\!(2-\sin^2{\theta})\frac{\cos{\kappa}}{\kappa} \\
&&V_{10}=\frac{1}{\sqrt{2}}\sin{2\theta}e^{i\varphi}\left[\frac{\cos{\kappa}}{\kappa} - 3\left(\frac{\sin{\kappa}}{\kappa^2} + \frac{\cos{\kappa}}{\kappa^3}\right)\right] \\
&&V_{00}=2\left[\left(1-3\cos^2{\theta}\right)\left(\frac{\sin{\kappa}}{\kappa^2} + \frac{\cos{\kappa}}{\kappa^3}\right)-\sin^2{\theta}\frac{\cos{\kappa}}{\kappa}\right]\\
&&V_{+-}=\sin^2{\theta}e^{2i\varphi}\left[3\left(\frac{\sin{\kappa}}{\kappa^2} + \frac{\cos{\kappa}}{\kappa^3}\right)-\frac{\cos{\kappa}}{\kappa}\right].
\end{eqnarray*}
All exchange rates between the internal states are proportional to the single-atom decay rate $\gamma$ and depend on the reduced distance between the two atoms $\kappa=2\pi r_{\alpha\beta}/\lambda$. Here, ${r}_{\alpha\beta}=\left|\mathbf{r}_{\alpha\beta}\right|$ is the modulus of the separation between the two atoms $\mathbf{r}_{\alpha\beta}=\mathbf{r}_\alpha-\mathbf{r}_\beta$, and $\theta$ and $\varphi$ are the angles between $\mathbf{r}_{\alpha\beta}$ and the transition dipole moment $\mathbf{d}^0$ and the $x$-axis, respectively [see Fig. \ref{fig:Fig1}(d)].

For small values of $\kappa$ (near-field) the exchange interactions (\ref{eq:V}) decay approximately as $1/\kappa^3$. Here, for a fixed value of $\kappa$, both the strength and sign of the interactions can be tuned by changing the angle $\theta$ [e.g. $V_{00}\approx 2\left(1-3\cos^2{\theta}\right)/\kappa^3$]. Control over this angle and, hence, the interactions, is obtained by applying a uniform magnetic field $\mathbf{B}=(B_x,B_y,B_z)$, represented in the master equation (\ref{equ:mastereq}) by substituting $H\to H+H_\Delta$, with
\begin{equation}
H_\Delta = \sum_{\alpha=1}^{N}\mathbf{d}_{\alpha}'\cdot\bar{\Delta}_{\alpha\alpha}\mathbf{d}_{\alpha}.
\label{equ:coupmat}
\end{equation}
Here, the matrix $\bar{\Delta}_{\alpha\alpha}$ reads
	\begin{equation}
		\bar{\Delta}_{\alpha\alpha} = \mu_\mathrm{B} g\begin{pmatrix}
B_z&\frac{B_x-iB_y}{\sqrt{2}}&0\\ 
\frac{B_x+iB_y}{\sqrt{2}} &0&\frac{B_x-iB_y}{\sqrt{2}}\\ 
0&\frac{B_x+iB_y}{\sqrt{2}}&-B_z\\
\end{pmatrix},\label{eq:Delta}
	\end{equation}
with $\mu_\mathrm{B}$ being the Bohr magneton and $g$ the Land\'e g-factor.

The second term of the master equation (\ref{equ:mastereq}) represents the dissipation via incoherent emission of photons into the radiation field and it is given by
	\begin{equation}
		\mathcal{D}(\rho) = \sum_{\alpha,\beta} \left(\mathbf{d}_{\alpha}\cdot\bar{\Gamma}_{\alpha\beta}\rho\mathbf{d}_{\beta}' - \frac{1}{2}\left\{\mathbf{d}_{\alpha}'\cdot\bar{\Gamma}_{\alpha\beta}\mathbf{d}_{\beta},\rho\right\}\right).
		\label{equ:dissipation}
	\end{equation} 	
The coefficient matrix $\bar{\Gamma}_{\alpha\beta}$ encodes the dissipative couplings between the atoms and has a similar structure to the coherent interaction matrix:
\begin{equation}
\bar{\Gamma}_{\alpha\beta} =\frac{3\gamma}{4} \begin{pmatrix}
\Gamma_{11}&-\Gamma_{10}^*&\Gamma_{+-}^*\\ 
-\Gamma_{10}&\Gamma_{00}&\Gamma_{10}^*\\ 
\Gamma_{+-}&\Gamma_{10}&\Gamma_{11}\\
\end{pmatrix}, \label{eq:G}
\end{equation}
with
\begin{eqnarray*}
&&\Gamma_{11} =(2-\sin^2{\theta})\frac{\sin{\kappa}}{\kappa}\! +\! \left(2-3\sin^2{\theta}\right)\left(\frac{\cos{\kappa}}{\kappa^2} - \frac{\sin{\kappa}}{\kappa^3}\right)\\
&&\Gamma_{10} = \frac{1}{\sqrt{2}}\sin{2\theta}e^{i\varphi}\left[-\frac{\sin{\kappa}}{\kappa} - 3\left(\frac{\cos{\kappa}}{\kappa^2} - \frac{\sin{\kappa}}{\kappa^3}\right)\right] \\
&&\Gamma_{00} = 2\left[\sin^2{\theta}\frac{\sin{\kappa}}{\kappa} + \left(1-3\cos^2{\theta}\right)\left(\frac{\cos{\kappa}}{\kappa^2} - \frac{\sin{\kappa}}{\kappa^3}\right)\right]\\
&&\Gamma_{+-} = \sin^2{\theta}e^{2i\varphi}\left[\frac{\sin{\kappa}}{\kappa} + 3\left(\frac{\cos{\kappa}}{\kappa^2} - \frac{\sin{\kappa}}{\kappa^3}\right)\right].
\end{eqnarray*}
The atoms couple to the radiation field as a collective and not as individuals. As a consequence, the decay rates in the system differ significantly from those of single emitters \cite{Dicke1954,Lehmberg1970,Agarwal1971}, with some being much larger and others much smaller than the single-atom decay rate $\gamma$ (corresponding to so-called superradiant and subradiant emission modes, respectively). This collective character becomes more pronounced for small reduced distances $\kappa$, i.e. small ratios $a/\lambda$, reaching regimes where some of the radiation modes are almost completely dark (with virtually zero decay rate). The population of these subradiant modes is the mechanism that allows for the prolongued storage of light in the atomic system.

\subsection{Dynamics in the single excitation sector}

Throughout, we will assume that the initial state contains a single excitation localized over a few lattice sites of the chain [Figs. \ref{fig:Fig1}(a) and (b)]. This single excitation (in one of the three states $\ket{-1}, \ket{0}$, or $\ket{+1}$) is transported via the exchange interactions given by $H$ (which conserve the number of excitations), while the action of dissipation can only decrease the number of excitations to zero. Thus, the dynamics can be described in a truncated space formed by the many-body ground state $\ket{G}\equiv\ket{g}_1\otimes\ket{g}_2\dots\otimes\ket{g}_N$ and the single-excitation states $\ket{e^m}_\alpha\equiv\ket{g}_1\otimes\ket{g}_2\dots\ket{m}_\alpha\dots\otimes\ket{g}_N$, for all $\alpha=1\dots N$ and $m=-1,0,+1$. Here, the density matrix takes the form
\begin{equation}
	\rho=\begin{pmatrix}
	\rho_{GG} & \bm{\rho}_{Ge}\\
	\bm{\rho}_{eG} & \bar{\rho}_{ee}
	\end{pmatrix},
	\label{equ:density}
\end{equation}
where $\rho_{GG} = \bra{G}\rho\ket{G}$, $\bm{\rho}_{Ge}=\bra{G}\rho\ket{\mathbf{e}}$, $\bm{\rho}_{eG}=\bra{\mathbf{e}}\rho\ket{G}$, and $\bar{\rho}_{ee}=\bra{\mathbf{e}}\rho\ket{\mathbf{e}}$, with $\ket{\mathbf{e}}$ being a row vector containing all single-excitation states $\ket{e^m}_\alpha$. The time-evolution of the elements of $\bar{\rho}_{ee}$ (a $3N\times 3N$ matrix) is decoupled from the dynamics of the remaining elements (see Appendix A), obeying the equation
\begin{equation}
\dot{\bar{\rho}}_{ee}  =   -\frac{\mi}{\hbar}\left[H_{\text{eff}},\bar{\rho}_{ee}\right],
\label{equ:masterred}
\end{equation}
with
\begin{equation}
H_\mathrm{eff}=\hbar\left(\bar{V}-\mi\frac{\bar{\Gamma}}{2}\right)+\bar{\Delta}.
\end{equation}
Here, $\bar{V}$, $\bar{\Gamma}$, and $\bar{\Delta}$, are $3N\times 3N$ matrices whose components for $\alpha,\beta=1,\dots,N$ are given by Eqs. (\ref{eq:V}), (\ref{eq:G}) and (\ref{eq:Delta}), respectively \footnote{Note that $\bar{V}_{\alpha\alpha}=0$ and $\bar{\Delta}_{\alpha\beta}=\bar{\Delta}_{\alpha\alpha}\delta_{\alpha\beta}$}.

We will consider in all cases a pure state as the initial state, and hence $\bar{\rho}_{ee}=\ket{\psi(t)}\bra{\psi(t)}$ with
\begin{equation}
	\ket{\psi(t)} = \sum_{m=\pm1,0}\sum_{\alpha=1}^N c_\alpha^{m}(t)\ket{e^m}_\alpha,
	\label{equ:3state}
\end{equation}
where $c_\alpha^{m}(t)$ is the probability amplitude of the $\alpha$-th atom being excited to the $\ket{m}$ state. The state (\ref{equ:3state}) evolves under the non-hermitian Hamiltonian $H_\mathrm{eff}$. The survival probability, i.e. the probability for not emitting a photon into the radiation field, is given by the norm of the wave function
\begin{equation}
	P_\mathrm{sur}(t) = \sum_{\alpha=1}^N \sum_{m = \pm1,0}|c_\alpha^{m}(t)|^2.
\end{equation}
The instantaneous photon emission rate, also called activity, is given by
\begin{equation}
  \left<K(t)\right>=\sum_{\alpha,\beta}\left<\mathbf{d}_{\alpha}'\cdot\bar{\Gamma}_{\alpha\beta}\mathbf{d}_{\beta}\right>.
\end{equation}
A value of the activity larger (smaller) than the single atom decay rate $\gamma$ for a state with large excitation density is indicative collective superradiant (subradiant) behavior of the photon emission.

\section{Subradiant transport on a ring lattice}

\begin{figure}[t!]
\begin{center}		
\includegraphics[width = \columnwidth]{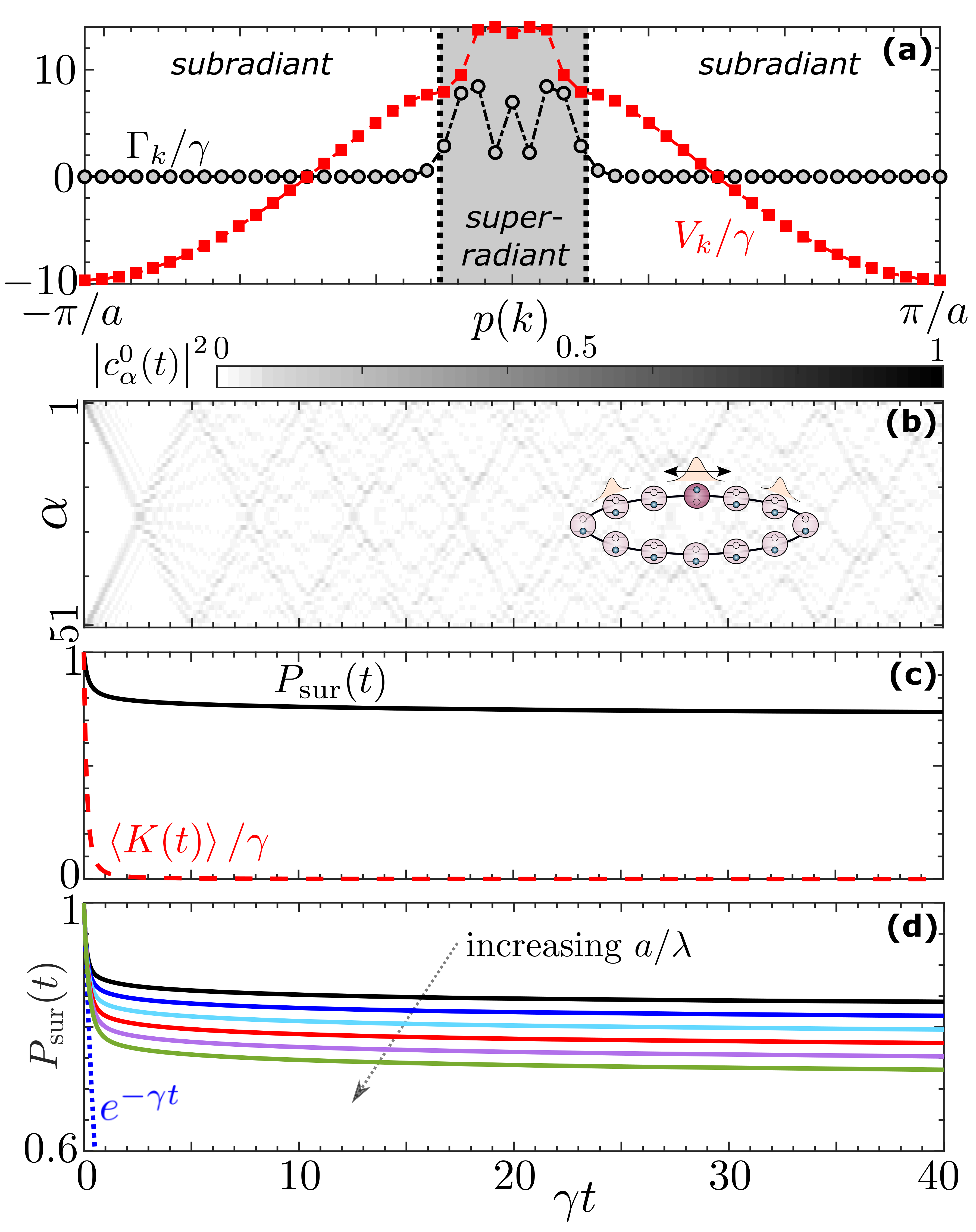}
\end{center}	
\caption{\textbf{Subradiance on a ring.} \textbf{(a):} Decay rates $\Gamma_k$ and energy $V_k$ of each mode $\ket{k_0}$ as a function of the momentum $p(k)=2 \pi k/(N a)$ for a for lattice formed by $N=51$ atoms and $a/\lambda=0.08$. The grey area indicates states which are superradiant, i.e. $\Gamma_k>\gamma$.  \textbf{(b):} Excitation probability $\left|c_\alpha^0(t)\right|^2$ at site $\alpha$ as a function of time given by (\ref{eq:ringtime}) for an initial excitation on site $\alpha=1$. \textbf{(c):} Probability of survival of the initial excitation $P_\mathrm{sur}$ (black solid line) and activity $\left<K\right>$ (red dashed line) as a function of time. \textbf{(d):} Survival probability as a function of time varying the ratio $a/\lambda$ from $0.05$ to $0.1$. The exponential decay of a single atom (dashed blue line) is shown for comparison.}
\label{fig:Fig2}
\end{figure}

First, we focus on a ring lattice, as illustrated in Fig. \ref{fig:Fig1}(a) \cite{Jen2018,Jen2018-1,Moreno2019}. Here, the matrices $\bar{V}$ and $\bar{\Gamma}$ are symmetric circulant due to the periodic boundary conditions such that, for each orientation of the transition dipole $m$, the simultaneous eigenstates for both matrices are given by the plane waves
\begin{equation}
\ket{k_m}=\frac{1}{\sqrt{N}}\sum_{\alpha=1}^N e^{i\frac{2\pi}{N}(\alpha-1)(k-q)}\ket{e^m}_\alpha,\label{eq:eigenring}
\end{equation}
with $k=-\left\lfloor{N/2}\right\rfloor,\dots \left\lfloor{(N-1)/2}\right\rfloor$ and $q=\left\lfloor{N/2}\right\rfloor$. For illustration purposes, we will consider in the following the case where only the $\ket{0}$ state is excited initially, and where the ring plane is perpendicular to the dipole moment $\mathbf{d}^0$ (i.e. $\theta=\pi/2$). Here, both the coherent and incoherent couplings between $\ket{0}$ and $\ket{\pm1}$ vanish [$V_{10}=0$ and $\Gamma_{10}=0$ in Eqs. (\ref{eq:V}) and (\ref{eq:G})], and hence the dynamics is determined solely by $V_{00}$ and $\Gamma_{00}$. The initial state can be written as
\begin{equation}
\ket{\psi(0)}=\sum_{\alpha} c_\alpha^0\ket{e^0}_\alpha=\frac{1}{\sqrt{N}}\sum_{k} c_k\ket{k_0}, \label{eq:gen_init}
\end{equation}
where the coefficients $c_\alpha^0$ and $c_k$ represent the probability amplitude distribution of the initial state in real and momentum space, respectively. Conversely, the time-evolved state takes the form
\begin{equation}
\ket{\psi(t)}=\frac{1}{\sqrt{N}}\sum_{k}c_ke^{-i\left(V_k-i\frac{\Gamma_k}{2}\right)t}\ket{k_0}, \label{eq:ringtime}
\end{equation}
where $V_k$ and $\Gamma_k$ are the eigenvalues of the matrices $\bar{V}$ and $\bar{\Gamma}$, respectively. Note that the collective character of the dissipation is reflected here in the decay rates $\Gamma_k$, which are either larger or smaller than the single atom decay rate $\gamma$, corresponding to $\ket{k_0}$ having superradiant or subradiant character, as illustrated in Fig. \ref{fig:Fig2}(a), while $V_k$ represents the energy of the corresponding mode.

Let us start by considering the initial state to be $\ket{e^0}_1$, i.e. an excitation localized on a single site of the lattice such that $c_\alpha^0=\delta_{\alpha1}$. This state can be written as a symmetric superposition of all plane waves (\ref{eq:eigenring}), i.e.,
\begin{equation}
\ket{\psi(0)}=\frac{1}{\sqrt{N}}\sum_{k}\ket{k_0}, \label{eq:init}
\end{equation}
and its time-evolution is then given by
\begin{equation}
\ket{\psi(t)}=\frac{1}{\sqrt{N}}\sum_{k}e^{-i\left(V_k-i\frac{\Gamma_k}{2}\right)t}\ket{k_0}. \label{eq:ringtime}
\end{equation}
This dynamics is depicted in Fig. \ref{fig:Fig2}(b), where we observe that the initial wave packet splits into two that travel in opposite directions. These wave packets disperse quickly due to the non-linearity of $V_k$ as a function of $p(k)=2 \pi k/(N a)$ [see Fig. \ref{fig:Fig2}(a)]. More importantly, the superradiant components (with large $\Gamma_k$) decay very fast and only the subradiant ones remain populated. This is seen in Fig. \ref{fig:Fig2}(c), which shows a plateau in the survival probability $P_\mathrm{sur}$, and near-zero emission rate $\left<K\right>$ after a rapid initial decay. The height of the plateau of $P_\mathrm{sur}$ is approximately given by the number of subradiant eigenstates [much larger than the number of superradiant ones, as can be seen in Fig. \ref{fig:Fig2}(a)] divided by the total number of modes $N$ \footnote{Note that this approximation only holds for small enough values of $a/\lambda$. As this ratio is increased, the decay rate $\Gamma_k$ of the subradiant states get closer to $\gamma$ and hence the plateau only holds for a small period of time $\tau$, given by the difference between $1/\Gamma_k$ and $1/\Gamma_{k+1}$.}. For a fixed value of $a/\lambda$, this ratio remains almost constant when increasing the number of atoms $N$. For a fixed size of the system $N$, on the other hand, reducing $a/\lambda$ increases the relative number of subradiant eigenstates and hence the survival probability, as can be seen in Fig. \ref{fig:Fig2}(d). In all cases considered, the lifetime of the excitation is dramatically longer than in the case of a single atom.

\begin{figure}[t!]
\begin{center}		
\includegraphics[width = \columnwidth]{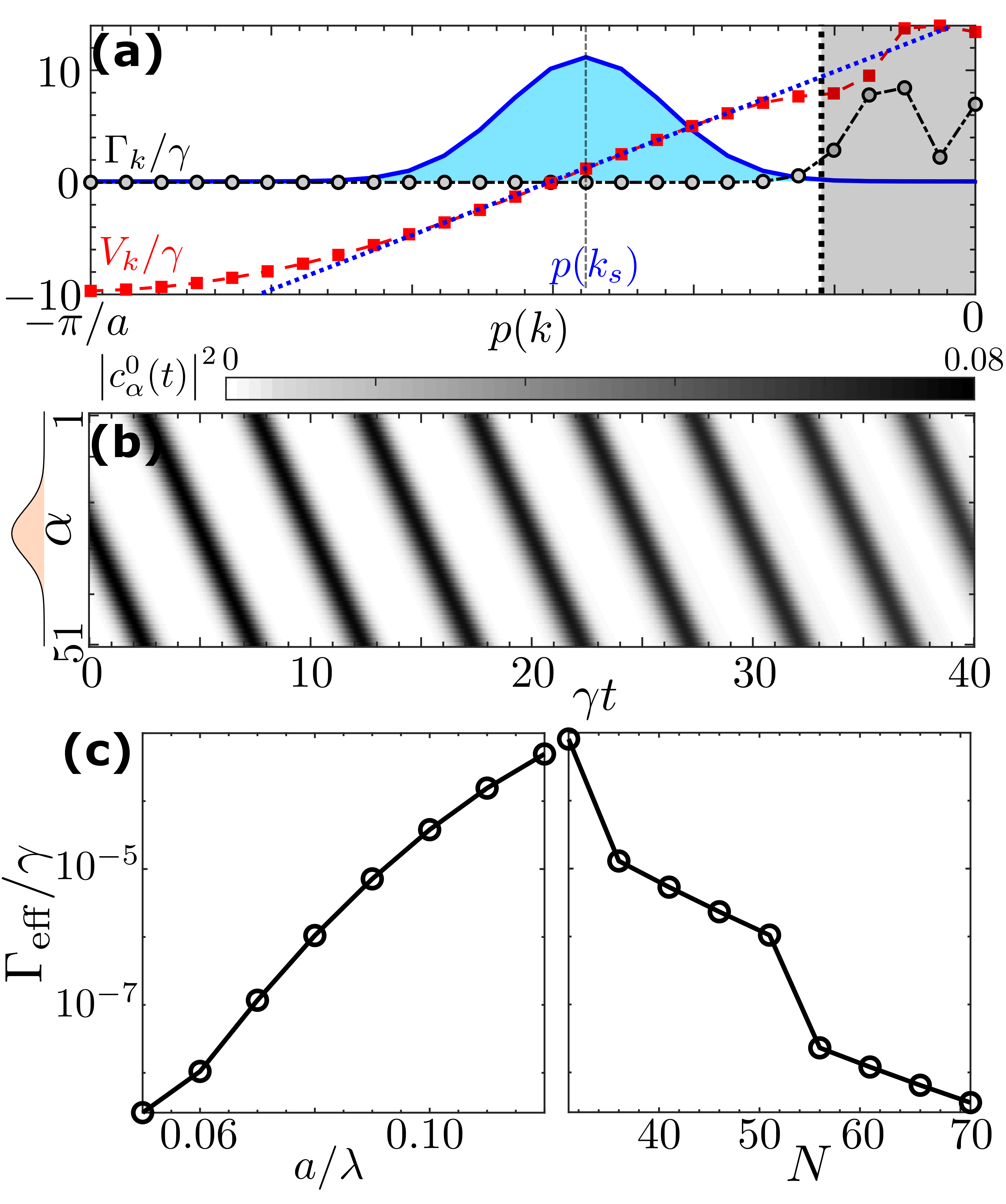}
\end{center}	
\caption{\textbf{Subradiance-protected wave packet.} \textbf{(a):} Decay rates $\Gamma_k$ and energy $V_k$ of each mode $\ket{k_0}$ as a function of the momentum $p(k)=2 \pi k/(N a)$ for a lattice formed by $N=51$ atoms and $a/\lambda=0.08$. The initial wave packet's probability distribution in momentum space (blue solid line) is centered at $p(k_s)\approx-0.43\pi/a$, with width $\sigma_k=\pi/(16 a)$. Here, the dispersion relation is approximately linear (blue dashed line to guide the eye). \textbf{(b):} Excitation probability $\left|c_\alpha^0(t)\right|^2$ at site $\alpha$ as a function of time. The initial wave packet's probability distribution in real space is sketched on the left. \textbf{(c):} Effective decay rate $\Gamma_\mathrm{eff}/\gamma$ of the excitation for $N=51$ as a function of $a/\lambda$ (left panel) and for $a/\lambda=0.08$ as a function of $N$ (right panel).}
\label{fig:Fig3}
\end{figure}

As can be observed in Figs. \ref{fig:Fig2}(a) and \ref{fig:Fig3}(a), the dispersion relation in the subradiant part of the spectrum is approximately linear [excluding the states with momentum $p(k)$ close to $\pm\pi/a$ and near the superradiant region]. Therefore, one can expect to have lossless-propagating wave packets with a constant group velocity (given by the gradient of $V_k$) without dispersing. We illustrate this by initialising the system with a Gaussian wave packet centered in momentum space at $p(k_s)$ (center of the linear dispersion manifold) and width $\sigma_k$ small enough to ensure that most components of the wave packet are located in the linear dispersion regime [see blue solid line in Fig. \ref{fig:Fig3}(a)]:
 \begin{equation}
\ket{\psi(0)}=\frac{1}{\sqrt{\sqrt{2\pi}\sigma_k}}\sum_{k} e^{-\frac{\left[p(k)-p(k_s)\right]^2}{4 \sigma_k^2}}\ket{k_0}. \label{eq:init_gau}
\end{equation}
In real space this is also a Gaussian wave packet
 \begin{equation}
\ket{\psi(0)}=\sqrt{\frac{\sigma_k}{\sqrt{2\pi}}}\sum_{\alpha} e^{-ip(k_s)a(\alpha-1)}e^{-\frac{\left[\alpha-1\right]^2a^2}{4/\sigma_k^2}}\ket{e^0}_\alpha, \label{eq:init_gau_real}
\end{equation}
whose probability distribution is sketched on the left hand side of Fig. \ref{fig:Fig3}(b). Here, it is shown that such wave packet travels indeed without appreciable dispersion around the ring. Moreover, the lifetime of the excitation is extremely long: its effective decay rate $\Gamma_\mathrm{eff}$ is six orders of magnitude smaller than the single atom decay rate $\gamma$. A similar reduction of the decay rate $\Gamma_\mathrm{eff}$ is also achieved for different system sizes $N$ and ratios $a/\lambda$, as it can be observed in Fig. \ref{fig:Fig3}(c).

\section{Finite linear chain: storage and transport control via magnetic field switching}

\begin{figure}[t!]
\begin{center}		
\includegraphics[width = \columnwidth]{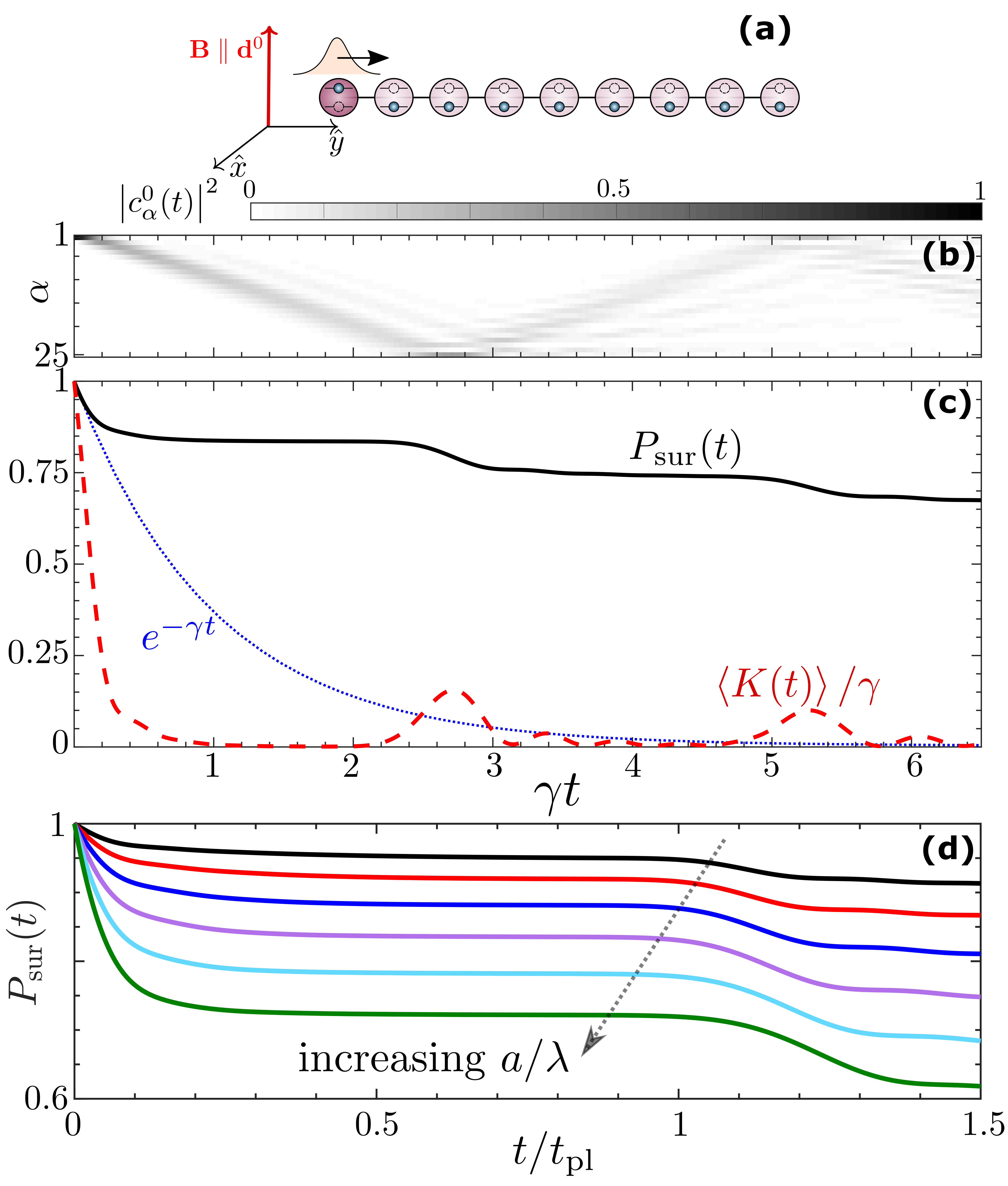}
\end{center}	
\caption{\textbf{Subradiance on a linear chain.} \textbf{(a):} Dynamics of a single excitation, initially on the leftmost atom ($\alpha=1$) with transition dipole moment perpendicular to the chain, propagating through the chain of $N=25$ atoms with $a/\lambda=0.08$. \textbf{(b):} Excitation probability \(|c^0_\alpha(t)|^2\) at each site $\alpha$ as a function of time. {\textbf{(c):}} Survival probability \(P_{\text{sur}}(t)\) (solid black) and activity \(\langle K(t)\rangle/\gamma\) (dashed red) as a function of time. The blue dotted line represents the survival probability in a non-interacting case. \textbf{(d):} Survival probability as a function of time, varying the ratio $a/\lambda$ from $0.05$ to $0.1$. The time is given in units of $t_\mathrm{pl}$ (see main text). }
\label{fig:Fig4}
\end{figure}

In this Section we will focus on the control of the subradiant excitation transport and storage on a linear one-dimensional chain with open boundary conditions, as depicted in Fig. \ref{fig:Fig4}(a) \cite{Filipp2011,Jen2016,Jen2017,Asenjo-Garcia2017}. We consider the initial state to be $\ket{e^0}_1$, representing one excitation at the leftmost site with the rest of the atoms in the ground state. We further assume that a uniform magnetic field $\mathbf{B}$ is applied perpendicularly to the chain (which lies on the $y$-axis) and parallel to $\mathbf{d}^0$, such that $\theta=\pi/2$.

The initial excitation is transported to the right of the chain via the dipole-dipole interactions [see Fig. \ref{fig:Fig4}(b)] until it reaches the other edge of the lattice and bounces back. As in the case of the ring, the excitation quickly disperses, and acquires a subradiant character when reaching the bulk of the chain [see Fig. \ref{fig:Fig4}(c)]. However, as the excitation reaches the other edge, the survival probability decays faster, accompanied by an increase of the activity. Analogously with the case of the ring, the height of the plateau in $P_\mathrm{sur}$ can be increased by reducing the ratio $a/\lambda$, as illustrated in Fig. \ref{fig:Fig4}(d). Here, in order to facilitate the comparison, the time is scaled by $t_\mathrm{pl}$, which is approximately the time that the excitation takes to reach the middle of the chain (inversely proportional to the nearest neighbor exchange rate).

\begin{figure}[t!]
	\begin{center}		
		\includegraphics[width =\columnwidth]{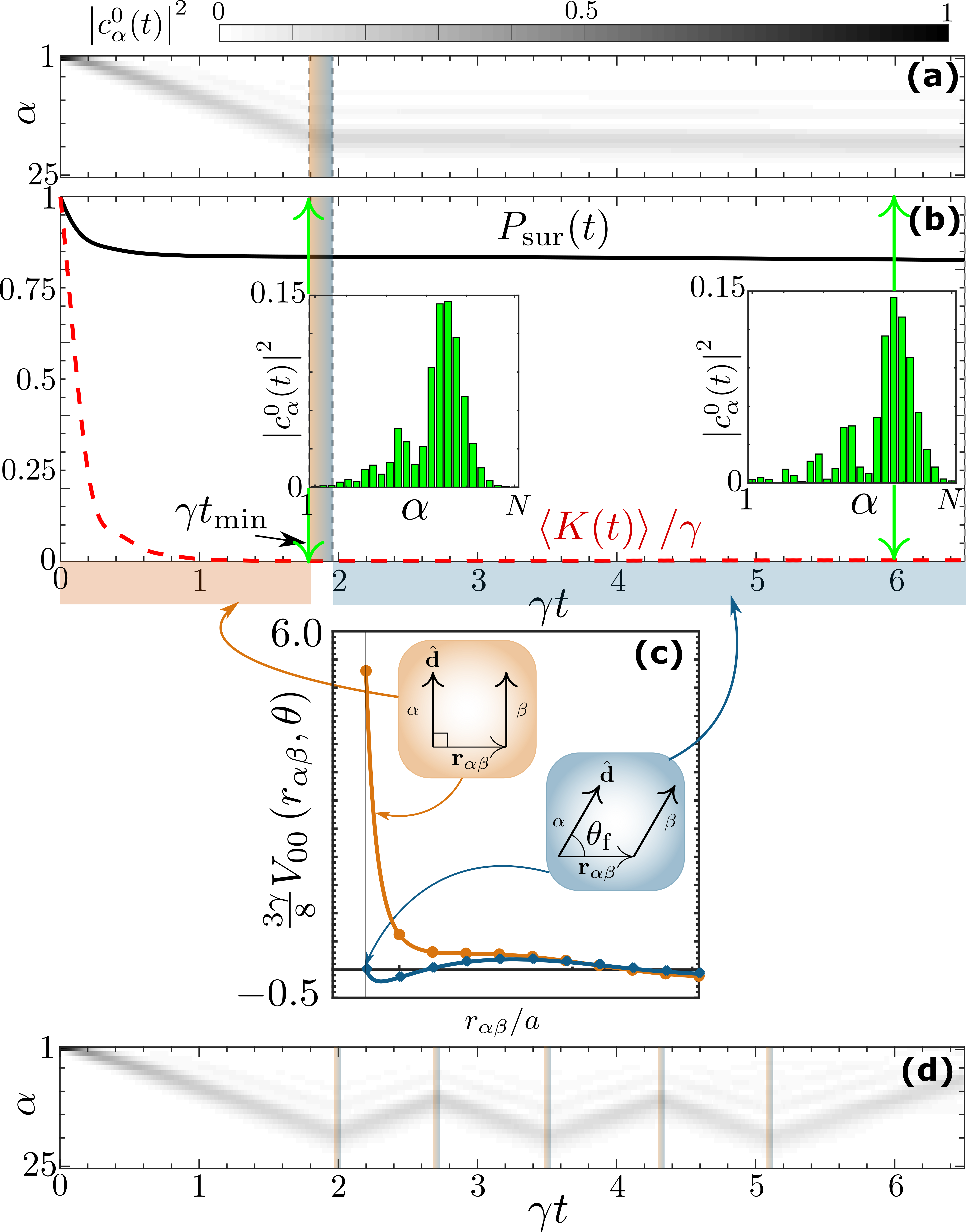}
	\end{center}
	\caption{\textbf{Excitation freezing.} Dynamics of a single excitation, initially under the same conditions as Fig. \ref{fig:Fig4}. At the time when the activity is minimised, $t_\mathrm{min}$, the magnetic field ($\Delta=10^3\gamma$) is rotated adiabatically (orange-blue shaded region between dashed lines) to the optimal angle for storage $\theta_\mathrm{f}$. \textbf{(a):} Excitation probability \(|c^0_\alpha(t)|^2\) at each site $\alpha$ as a function of time. {\textbf{(b):}} Survival probability $P_{\text{sur}}(t)$ (solid black) and activity $\langle K(t)\rangle/\gamma$ (dashed red) as a function of time. The insets show individual atom populations at $\gamma t_\mathrm{min} = 1.79$ and $\gamma t = 6$. \textbf{(c):} Interaction strengths in the 1D chain between two atoms in the $\ket{0}$ state: at $t=0$, with the dipole moments of the atoms aligned perpendicular to the atom separation (orange) and after the change of direction of the magnetic field, when the dipole moments have followed the change adiabatically and are aligned at an angle $\theta_\mathrm{f}$ in order to minimise the coupling between nearest neighbors (blue). \textbf{(d):} Excitation probability \(|c^0_\alpha(t)|^2\) at each site $\alpha$ as a function of time, where the magnetic field is repeatedly switched from $\pi/2$ to an angle where the nearest neighbor interactions change sign and viceversa.}
	\label{fig:Fig5}
\end{figure}

Since the excitation has almost zero decay rate within the bulk of the chain, its lifetime is ultimately limited by the time it takes for it to reach the other boundary, i.e. by size of the system and the value of the exchange interactions. One can ask, thus, whether it is possible to freeze the transport of the wave packet and confine it in the subradiant states of the bulk. This can indeed be done by adiabatically changing the direction of the external magnetic field, exploiting that the strength and sign of the exchange interactions depends on the angle between the transition dipole moment $\mathbf{d}^0$ and the direction of the separation between the atoms $\theta$.

Let us illustrate this protocol via an example depicted in Fig. \ref{fig:Fig5} for the same parameters used in Fig. \ref{fig:Fig4}. At $t=0$, $\theta=\pi/2$, such that the nearest neighbor interactions are larger than the single atom decay rate $\gamma$ [as shown in Fig. \ref{fig:Fig5}(c), orange line], and make the excitation propagate into the bulk [see Figs. \ref{fig:Fig5}(a) and (b)]. When the activity reaches a minimum at $t=t_\mathrm{min}$, the direction of the magnetic field is changed within the $yz$-plane. This switch is done adiabatically, such that it is followed by the transition dipole moment of the excitation (i.e. the switching time $\tau\gg 1/\Delta$), but quickly enough to keep the excitation from leaving the bulk of the lattice. The change of the magnetic field direction is mathematically equivalent to a rotation of the angle $\theta$ between the quantization axis and the chain from its initial value $\theta_\mathrm{in}=\pi/2$ to a final value $\theta_\mathrm{f}$, which leads in turn to modified interactions. In particular, in order to slow down the excitation transport in the bulk, we fix the final value such that the nearest-neighbor interaction coefficient is zero, $V_{00}(a,\theta_\mathrm{f})=0$ [see Fig. \ref{fig:Fig5}(c), blue line]. While this change does not freeze the excitation transport entirely due to the non-zero values of the exchange rates beyond nearest neighbors, it does slow it down notably, as one can see in Fig. \ref{fig:Fig5}(a). Most importantly, the subradiant character of the propagation is preserved, reflected in a constant survival probability $P_\mathrm{sur}$ and vanishing activity, as shown in Fig. \ref{fig:Fig5}(b).

The versatility of the system using the change in the magnetic field direction is further illustrated in Fig. \ref{fig:Fig5}(d). Here, we show an example where several changes in the direction of the magnetic field allow to switch the direction of travel of the excitation. Most importantly, the activity remains close to zero throughout all of these changes, as long as the excitation stays in the bulk of the chain.

\begin{figure}[t!]
\begin{center}		
\includegraphics[width = \linewidth]{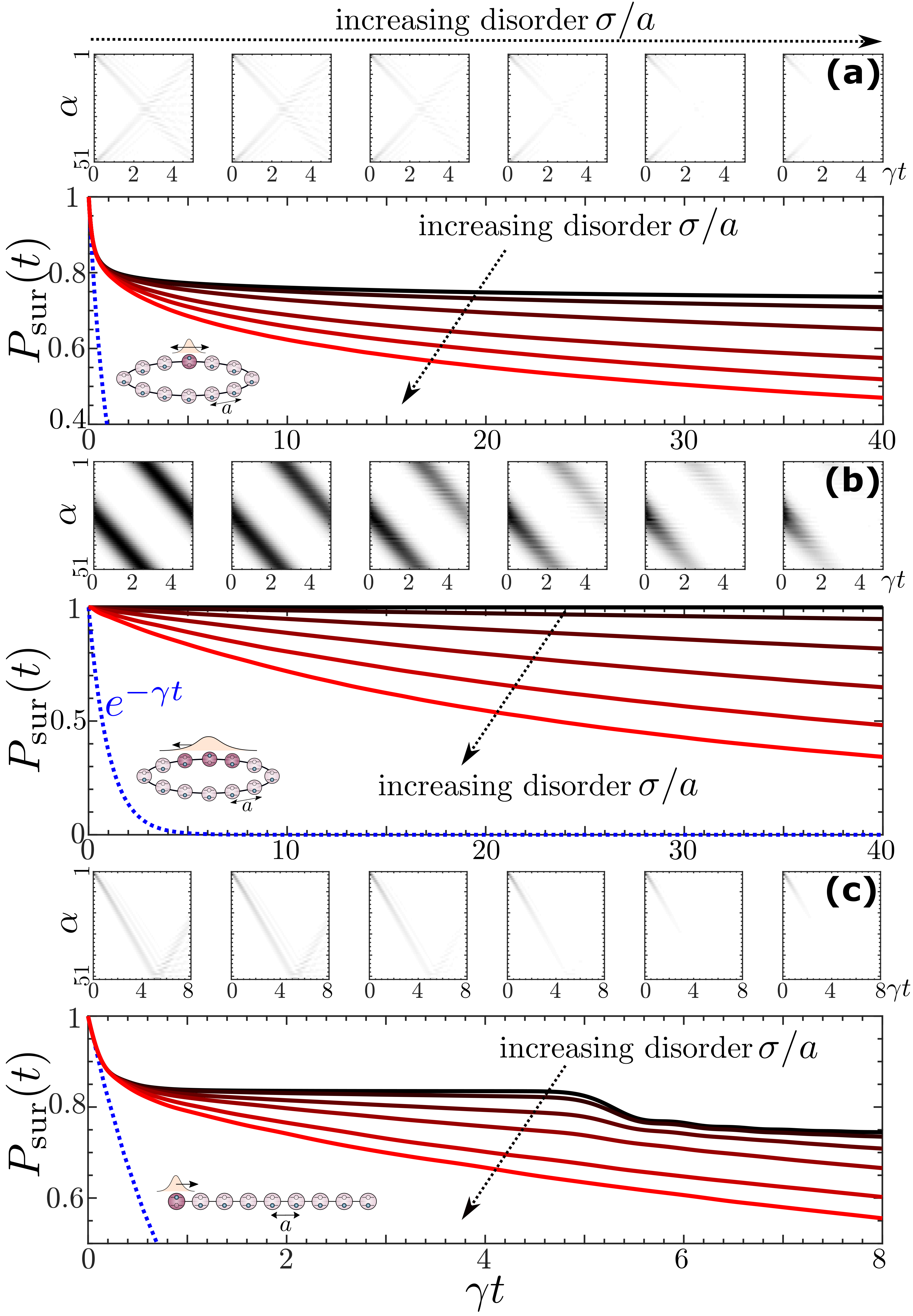}
\end{center}	
\caption{\textbf{Disorder.} Survival probability for an initial excitation on a chain of $N=51$ atoms and $a/\lambda=0.08$ \textbf{(a):} on a single site on a ring lattice, \textbf{(b):} on an extended gaussian wave packet on a ring lattice and \textbf{(c):} on a single site on a linear 1D lattice. All panels show the excitation probability $\left|c_\alpha^0(t)\right|^2$ at site $\alpha$ as a function of time and the Survival probability without disorder (leftmost panel, black solid line) and in the presence of disorder (rest of panels and solid red lines), modelled by a gaussian distribution of the positions of the atoms around the center of each site with widths $\sigma=0.01a\to0.05 a$ (average over 100 iterations of the disorder). The blue dashed line in all cases represents an exponential decay with single atom decay rate $\gamma$ for comparison.}
\label{fig:Fig6}
\end{figure}

\section{Disorder}

Finally, we briefly consider the effect of disorder on the subradiant transport discussed in the previous sections. In particular, we consider the disorder introduced due to the finite width of the external wavefunction of each atom, which we model as a three dimensional Gaussian with width $\sigma$ centered in the respective lattice sites.

Since the long-ranged exchange interactions $\bar{V}$ [given by (\ref{eq:V})] are functions of the separation between the atoms, the uncertainty in the atomic positions translates into disorder in the hopping rates in the exchange Hamiltonian $H$ given by Eq.  (\ref{equ:hamiltonian}). This kind of positional disorder in Hamiltonians with long-ranged hopping has been recently studied and found to give rise to localization \cite{Deng2018}. Consistently with this, we find that as we increase $\sigma$ the eigenstates of the Hamiltonian become localized, inhibiting transport. This  can be seen in the top panels of Figs. \ref{fig:Fig6}(a) (b) and (c), where we show the excitation probability $\left|c_\alpha^0(t)\right|^2$ as a function of time for increasing disorder (ratio $\sigma/a$) from left to right.

Subradiance is, however, not a fine-tuned property but rather known to be robust in the presence of disorder \cite{Akkermans2013,Biella2013}. Indeed, in each lower panel in Figs. \ref{fig:Fig6}(a), (b) and (c) one can observe that, while increasing disorder has a detrimental effect on the subradiant state manifold, in all cases considered the excitation features lifetimes dramatically longer than the ones of an individual atom.

\section{Conclusions and outlook}

We have investigated the transport of an excitation in a one-dimensional atomic lattice that occurs due to the coupling of the atoms to the radiation field. In particular, we have shown that there is a high dimensional subradiant manifold that allows for dispersionless transport, control and storage of wave packets.

However, there are a number of experimental challenges to overcome when considering the realization of such long lived excitation storage in a dissipative system. One is to achieve a sufficiently small ratio $a/\lambda$ such that highly subradiant states emerge. An example of such a system, using a transition in the triplet series of alkaline earth metal atoms with a particularly long wavelength ($2.6$ $\mu$m in strontium), was introduced in \cite{Olmos2013}. The trapping of these alkaline earth metal atoms is currently realized experimentally both in optical lattices \cite{Yamamoto2016,Miranda2017} and tweezer arrays \cite{Norcia2018}. Even smaller ratios $a/\lambda$ can be achieved by using Rydberg states, where the transition wavelengths are much longer than in low-lying states. An alternative approach that allows subradiant states to emerge for large ratios $a/\lambda$ is changing the radiation field's boundary conditions by placing, e.g., a surface \cite{Jones2018} or a waveguide \cite{Asenjo-Garcia2017,LeKien2017,Lodahl2017,Buonaiuto2019,Albrecht2019} near the atoms, which in turn modifies the exchange interaction and dissipation. Another experimental challenge is the preparation of the subradiant wave packets. In particular, preparing states with one excitation localized on one or a few sites will require single-site resolution and addressability, which has been achieved experimentally in optical lattices and tweezer arrays \cite{Weitenberg2011,Kim2016,Endres2016,Barredo2016,Cooper2018}. Moreover, creating a wave packet with a linear dispersion relation will require a phase imprinting mechanism, which may be challenging to implement experimentally.

A future direction connecting to this work will be to move away from the linear optics regime (single excitation sector of the dynamics) \cite{Asenjo-Garcia2017,Jen2018,Moreno2019,Guimond2019,Zhang2019} and consider situations where two or more wave packets interfere with each other effectively realizing photon-photon interactions in a subradiant decoherence-free manifold. Such platform can find applications ranging from the creation of non-classical states of light to the realization of photon-photon quantum gates \cite{Gorshkov2011,Tiarks2019}.

\section*{acknowledgements}

The authors acknowledge fruitful discussions with Katarzyna Macieszczak, Matteo Marcuzzi, and Thomas Fernholz. BO was supported by the Royal Society and EPSRC [Grant No. DH130145]. The research leading to these results has received funding from the European Research Council under the European Union's Seventh Framework Programme (FP/2007-2013)  [ERC Grant Agreement No. 335266 (ESCQUMA)] and from the European Union's H2020 research and innovation programme [Grant Agreement No. 800942 (ErBeStA)]. Funding was also received from the EPSRC [Grant No. EP/M014266/1 and No. EP/R04340X/1]. IL gratefully acknowledges funding through the Royal Society Wolfson Research Merit Award.

\appendix	

\section{Dynamics in the truncated Hilbert space}\label{app:rhoeqs}

In this Appendix we give the expressions for the equations of motion of each component of the truncated density matrix (\ref{equ:density}). We obtain these equations by simply substituting the expression (\ref{equ:density}) into the master equation (\ref{equ:mastereq}), such that we can split the density matrix into the four components $\rho_{GG}=\bra{G}\rho\ket{G}$, $\rho_{Ge^m_\alpha}=\bra{G}\rho\ket{e^m}_\alpha$, $\rho_{e^m_\alpha G}={}_\alpha\bra{e^m}\rho\ket{G}$ and $\rho_{e^m_\alpha e^l_\beta}={}_\alpha\bra{e^m}\rho\ket{e^l}_\beta$. The time evolution of each component is given by
\begin{equation}
\begin{aligned}
\dot{\rho}_{GG} = & \sum_{\gamma\epsilon}\sum_{np} \Gamma_{\gamma\epsilon}^{np}\rho_{e^{n}_{\gamma}e^{p}_{\epsilon}} \\
\dot{\rho}_{Ge^{m}_{\alpha}} = & \frac{\mi}{\hbar}\sum_{\gamma}\sum_{n}\rho_{Ge_{\gamma}^n}\left[\hbar Z^{nm*}_{\gamma\alpha}+\Delta_{\gamma\gamma}^{nm}\delta_{\gamma\alpha}\right] \\
\dot{\rho}_{e^{m}_{\alpha}G} = &  - \frac{\mi}{\hbar}\sum_{\gamma}\sum_{n}\left[\hbar Z^{mn}_{\alpha\gamma}+\Delta_{\gamma\gamma}^{mn}\delta_{\gamma\alpha}\right]\rho_{e_{\gamma}^nG}
\end{aligned}
\end{equation}
and 
\begin{equation}
\begin{aligned}
\dot{\rho}_{e^{m}_{\alpha}e^{l}_{\beta}} = &  - \frac{\mi}{\hbar}\sum_{\gamma}\sum_{n}\left[\left(\hbar Z_{\alpha\gamma}^{mn}+\Delta_{\gamma\gamma}^{mn}\delta_{\gamma\alpha}\right)\rho_{e_\gamma^ne^{l}_{\beta}}\right.\\
& \left.- \rho_{e_{\alpha}^me_{\gamma}^n}\left(\hbar Z^{nl*}_{\gamma\beta}+\Delta_{\gamma\gamma}^{nl}\delta_{\gamma\beta}\right)\right],
\end{aligned}
\end{equation}
where $Z_{\alpha\beta}^{ml}=V_{\alpha\beta}^{ml}-\frac{\mi}{2}\Gamma_{\alpha\beta}^{ml}$ [Eqs. (\ref{eq:V}) and (\ref{eq:G})].


\begin{thebibliography}{66}%
\makeatletter
\providecommand \@ifxundefined [1]{%
 \@ifx{#1\undefined}
}%
\providecommand \@ifnum [1]{%
 \ifnum #1\expandafter \@firstoftwo
 \else \expandafter \@secondoftwo
 \fi
}%
\providecommand \@ifx [1]{%
 \ifx #1\expandafter \@firstoftwo
 \else \expandafter \@secondoftwo
 \fi
}%
\providecommand \natexlab [1]{#1}%
\providecommand \enquote  [1]{``#1''}%
\providecommand \bibnamefont  [1]{#1}%
\providecommand \bibfnamefont [1]{#1}%
\providecommand \citenamefont [1]{#1}%
\providecommand \href@noop [0]{\@secondoftwo}%
\providecommand \href [0]{\begingroup \@sanitize@url \@href}%
\providecommand \@href[1]{\@@startlink{#1}\@@href}%
\providecommand \@@href[1]{\endgroup#1\@@endlink}%
\providecommand \@sanitize@url [0]{\catcode `\\12\catcode `\$12\catcode
  `\&12\catcode `\#12\catcode `\^12\catcode `\_12\catcode `\%12\relax}%
\providecommand \@@startlink[1]{}%
\providecommand \@@endlink[0]{}%
\providecommand \url  [0]{\begingroup\@sanitize@url \@url }%
\providecommand \@url [1]{\endgroup\@href {#1}{\urlprefix }}%
\providecommand \urlprefix  [0]{URL }%
\providecommand \Eprint [0]{\href }%
\providecommand \doibase [0]{http://dx.doi.org/}%
\providecommand \selectlanguage [0]{\@gobble}%
\providecommand \bibinfo  [0]{\@secondoftwo}%
\providecommand \bibfield  [0]{\@secondoftwo}%
\providecommand \translation [1]{[#1]}%
\providecommand \BibitemOpen [0]{}%
\providecommand \bibitemStop [0]{}%
\providecommand \bibitemNoStop [0]{.\EOS\space}%
\providecommand \EOS [0]{\spacefactor3000\relax}%
\providecommand \BibitemShut  [1]{\csname bibitem#1\endcsname}%
\let\auto@bib@innerbib\@empty
\bibitem [{\citenamefont {Dicke}(1954)}]{Dicke1954}%
  \BibitemOpen
  \bibfield  {author} {\bibinfo {author} {\bibfnamefont {R.~H.}\ \bibnamefont
  {Dicke}},\ }\bibfield  {title} {\enquote {\bibinfo {title} {{Coherence in
  spontaneous radiation processes}},}\ }\href {\doibase 10.1103/PhysRev.93.99}
  {\bibfield  {journal} {\bibinfo  {journal} {Phys. Rev.}\ }\textbf {\bibinfo
  {volume} {93}},\ \bibinfo {pages} {99} (\bibinfo {year} {1954})},\ \Eprint
  {http://arxiv.org/abs/1407.7336} {arXiv:1407.7336} \BibitemShut {NoStop}%
\bibitem [{\citenamefont {Lehmberg}(1970)}]{Lehmberg1970}%
  \BibitemOpen
  \bibfield  {author} {\bibinfo {author} {\bibfnamefont {R.~H.}\ \bibnamefont
  {Lehmberg}},\ }\bibfield  {title} {\enquote {\bibinfo {title} {{Radiation
  from an N-atom system. I. General formalism}},}\ }\href {\doibase
  10.1103/PhysRevA.2.883} {\bibfield  {journal} {\bibinfo  {journal} {Phys.
  Rev. A}\ }\textbf {\bibinfo {volume} {2}},\ \bibinfo {pages} {883} (\bibinfo
  {year} {1970})}\BibitemShut {NoStop}%
\bibitem [{\citenamefont {Agarwal}(1971)}]{Agarwal1971}%
  \BibitemOpen
  \bibfield  {author} {\bibinfo {author} {\bibfnamefont {G.~S.}\ \bibnamefont
  {Agarwal}},\ }\bibfield  {title} {\enquote {\bibinfo {title}
  {{Master-equation approach to spontaneous emission. II. Emission from a
  system of harmonic oscillators}},}\ }\href {\doibase 10.1103/PhysRevA.3.1783}
  {\bibfield  {journal} {\bibinfo  {journal} {Phys. Rev. A}\ }\textbf {\bibinfo
  {volume} {3}},\ \bibinfo {pages} {1783} (\bibinfo {year} {1971})}\BibitemShut
  {NoStop}%
\bibitem [{\citenamefont {Olmos}\ \emph {et~al.}(2013)\citenamefont {Olmos},
  \citenamefont {Yu}, \citenamefont {Singh}, \citenamefont {Schreck},
  \citenamefont {Bongs},\ and\ \citenamefont {Lesanovsky}}]{Olmos2013}%
  \BibitemOpen
  \bibfield  {author} {\bibinfo {author} {\bibfnamefont {B.}~\bibnamefont
  {Olmos}}, \bibinfo {author} {\bibfnamefont {D.}~\bibnamefont {Yu}}, \bibinfo
  {author} {\bibfnamefont {Y.}~\bibnamefont {Singh}}, \bibinfo {author}
  {\bibfnamefont {F.}~\bibnamefont {Schreck}}, \bibinfo {author} {\bibfnamefont
  {K.}~\bibnamefont {Bongs}}, \ and\ \bibinfo {author} {\bibfnamefont
  {I.}~\bibnamefont {Lesanovsky}},\ }\bibfield  {title} {\enquote {\bibinfo
  {title} {{Long-range interacting many-body systems with alkaline-earth-metal
  atoms}},}\ }\href {\doibase 10.1103/PhysRevLett.110.143602} {\bibfield
  {journal} {\bibinfo  {journal} {Phys. Rev. Lett.}\ }\textbf {\bibinfo
  {volume} {110}},\ \bibinfo {pages} {143602} (\bibinfo {year} {2013})},\
  \Eprint {http://arxiv.org/abs/1211.4537} {arXiv:1211.4537} \BibitemShut
  {NoStop}%
\bibitem [{\citenamefont {Sutherland}\ and\ \citenamefont
  {Robicheaux}(2016)}]{Sutherland2016}%
  \BibitemOpen
  \bibfield  {author} {\bibinfo {author} {\bibfnamefont {R.~T.}\ \bibnamefont
  {Sutherland}}\ and\ \bibinfo {author} {\bibfnamefont {F.}~\bibnamefont
  {Robicheaux}},\ }\bibfield  {title} {\enquote {\bibinfo {title} {Collective
  dipole-dipole interactions in an atomic array},}\ }\href {\doibase
  10.1103/PhysRevA.94.013847} {\bibfield  {journal} {\bibinfo  {journal} {Phys.
  Rev. A}\ }\textbf {\bibinfo {volume} {94}},\ \bibinfo {pages} {013847}
  (\bibinfo {year} {2016})}\BibitemShut {NoStop}%
\bibitem [{\citenamefont {Bettles}\ \emph {et~al.}(2016)\citenamefont
  {Bettles}, \citenamefont {Gardiner},\ and\ \citenamefont
  {Adams}}]{Bettles2016}%
  \BibitemOpen
  \bibfield  {author} {\bibinfo {author} {\bibfnamefont {Robert~J.}\
  \bibnamefont {Bettles}}, \bibinfo {author} {\bibfnamefont {Simon~A.}\
  \bibnamefont {Gardiner}}, \ and\ \bibinfo {author} {\bibfnamefont
  {Charles~S.}\ \bibnamefont {Adams}},\ }\bibfield  {title} {\enquote {\bibinfo
  {title} {Cooperative eigenmodes and scattering in one-dimensional atomic
  arrays},}\ }\href {\doibase 10.1103/PhysRevA.94.043844} {\bibfield  {journal}
  {\bibinfo  {journal} {Phys. Rev. A}\ }\textbf {\bibinfo {volume} {94}},\
  \bibinfo {pages} {043844} (\bibinfo {year} {2016})}\BibitemShut {NoStop}%
\bibitem [{\citenamefont {Friedberg}\ \emph {et~al.}(1973)\citenamefont
  {Friedberg}, \citenamefont {Hartmann},\ and\ \citenamefont
  {Manassah}}]{Friedberg1973}%
  \BibitemOpen
  \bibfield  {author} {\bibinfo {author} {\bibfnamefont {R.}~\bibnamefont
  {Friedberg}}, \bibinfo {author} {\bibfnamefont {S.R.}\ \bibnamefont
  {Hartmann}}, \ and\ \bibinfo {author} {\bibfnamefont {J.T.}\ \bibnamefont
  {Manassah}},\ }\bibfield  {title} {\enquote {\bibinfo {title} {Frequency
  shifts in emission and absorption by resonant systems ot two-level atoms},}\
  }\href {\doibase https://doi.org/10.1016/0370-1573(73)90001-X} {\bibfield
  {journal} {\bibinfo  {journal} {Phys. Rep.}\ }\textbf {\bibinfo {volume}
  {7}},\ \bibinfo {pages} {101} (\bibinfo {year} {1973})}\BibitemShut {NoStop}%
\bibitem [{\citenamefont {Scully}(2009)}]{Scully2009}%
  \BibitemOpen
  \bibfield  {author} {\bibinfo {author} {\bibfnamefont {Marlan~O.}\
  \bibnamefont {Scully}},\ }\bibfield  {title} {\enquote {\bibinfo {title}
  {Collective lamb shift in single photon dicke superradiance},}\ }\href
  {\doibase 10.1103/PhysRevLett.102.143601} {\bibfield  {journal} {\bibinfo
  {journal} {Phys. Rev. Lett.}\ }\textbf {\bibinfo {volume} {102}},\ \bibinfo
  {pages} {143601} (\bibinfo {year} {2009})}\BibitemShut {NoStop}%
\bibitem [{\citenamefont {Keaveney}\ \emph {et~al.}(2012)\citenamefont
  {Keaveney}, \citenamefont {Sargsyan}, \citenamefont {Krohn}, \citenamefont
  {Hughes}, \citenamefont {Sarkisyan},\ and\ \citenamefont
  {Adams}}]{Keaveney2012}%
  \BibitemOpen
  \bibfield  {author} {\bibinfo {author} {\bibfnamefont {J.}~\bibnamefont
  {Keaveney}}, \bibinfo {author} {\bibfnamefont {A.}~\bibnamefont {Sargsyan}},
  \bibinfo {author} {\bibfnamefont {U.}~\bibnamefont {Krohn}}, \bibinfo
  {author} {\bibfnamefont {I.~G.}\ \bibnamefont {Hughes}}, \bibinfo {author}
  {\bibfnamefont {D.}~\bibnamefont {Sarkisyan}}, \ and\ \bibinfo {author}
  {\bibfnamefont {C.~S.}\ \bibnamefont {Adams}},\ }\bibfield  {title} {\enquote
  {\bibinfo {title} {{Cooperative Lamb shift in an atomic vapor layer of
  nanometer thickness}},}\ }\href {\doibase 10.1103/PhysRevLett.108.173601}
  {\bibfield  {journal} {\bibinfo  {journal} {Phys. Rev. Lett.}\ }\textbf
  {\bibinfo {volume} {108}},\ \bibinfo {pages} {173601} (\bibinfo {year}
  {2012})},\ \Eprint {http://arxiv.org/abs/1201.5251} {arXiv:1201.5251}
  \BibitemShut {NoStop}%
\bibitem [{\citenamefont {Pellegrino}\ \emph {et~al.}(2014)\citenamefont
  {Pellegrino}, \citenamefont {Bourgain}, \citenamefont {Jennewein},
  \citenamefont {Sortais}, \citenamefont {Browaeys}, \citenamefont {Jenkins},\
  and\ \citenamefont {Ruostekoski}}]{Pellegrino2014}%
  \BibitemOpen
  \bibfield  {author} {\bibinfo {author} {\bibfnamefont {J.}~\bibnamefont
  {Pellegrino}}, \bibinfo {author} {\bibfnamefont {R.}~\bibnamefont
  {Bourgain}}, \bibinfo {author} {\bibfnamefont {S.}~\bibnamefont {Jennewein}},
  \bibinfo {author} {\bibfnamefont {Y.~R.P.}\ \bibnamefont {Sortais}}, \bibinfo
  {author} {\bibfnamefont {A.}~\bibnamefont {Browaeys}}, \bibinfo {author}
  {\bibfnamefont {S.~D.}\ \bibnamefont {Jenkins}}, \ and\ \bibinfo {author}
  {\bibfnamefont {J.}~\bibnamefont {Ruostekoski}},\ }\bibfield  {title}
  {\enquote {\bibinfo {title} {{Observation of suppression of light scattering
  induced by dipole-dipole interactions in a cold-atom ensemble}},}\ }\href
  {\doibase 10.1103/PhysRevLett.113.133602} {\bibfield  {journal} {\bibinfo
  {journal} {Phys. Rev. Lett.}\ }\textbf {\bibinfo {volume} {113}},\ \bibinfo
  {pages} {133602} (\bibinfo {year} {2014})},\ \Eprint
  {http://arxiv.org/abs/1402.4167} {arXiv:1402.4167} \BibitemShut {NoStop}%
\bibitem [{\citenamefont {Jenkins}\ \emph {et~al.}(2016)\citenamefont
  {Jenkins}, \citenamefont {Ruostekoski}, \citenamefont {Javanainen},
  \citenamefont {Jennewein}, \citenamefont {Bourgain}, \citenamefont
  {Pellegrino}, \citenamefont {Sortais},\ and\ \citenamefont
  {Browaeys}}]{Jenkins2016}%
  \BibitemOpen
  \bibfield  {author} {\bibinfo {author} {\bibfnamefont {S.~D.}\ \bibnamefont
  {Jenkins}}, \bibinfo {author} {\bibfnamefont {J.}~\bibnamefont
  {Ruostekoski}}, \bibinfo {author} {\bibfnamefont {J.}~\bibnamefont
  {Javanainen}}, \bibinfo {author} {\bibfnamefont {S.}~\bibnamefont
  {Jennewein}}, \bibinfo {author} {\bibfnamefont {R.}~\bibnamefont {Bourgain}},
  \bibinfo {author} {\bibfnamefont {J.}~\bibnamefont {Pellegrino}}, \bibinfo
  {author} {\bibfnamefont {Y.~R.~P.}\ \bibnamefont {Sortais}}, \ and\ \bibinfo
  {author} {\bibfnamefont {A.}~\bibnamefont {Browaeys}},\ }\bibfield  {title}
  {\enquote {\bibinfo {title} {Collective resonance fluorescence in small and
  dense atom clouds: Comparison between theory and experiment},}\ }\href
  {\doibase 10.1103/PhysRevA.94.023842} {\bibfield  {journal} {\bibinfo
  {journal} {Phys. Rev. A}\ }\textbf {\bibinfo {volume} {94}},\ \bibinfo
  {pages} {023842} (\bibinfo {year} {2016})}\BibitemShut {NoStop}%
\bibitem [{\citenamefont {Bromley}\ \emph {et~al.}(2016)\citenamefont
  {Bromley}, \citenamefont {Zhu}, \citenamefont {Bishof}, \citenamefont
  {Zhang}, \citenamefont {Bothwell}, \citenamefont {Schachenmayer},
  \citenamefont {Nicholson}, \citenamefont {Kaiser}, \citenamefont {Yelin},
  \citenamefont {Lukin}, \citenamefont {Rey},\ and\ \citenamefont
  {Ye}}]{Bromley2016}%
  \BibitemOpen
  \bibfield  {author} {\bibinfo {author} {\bibfnamefont {S.~L.}\ \bibnamefont
  {Bromley}}, \bibinfo {author} {\bibfnamefont {B.}~\bibnamefont {Zhu}},
  \bibinfo {author} {\bibfnamefont {M.}~\bibnamefont {Bishof}}, \bibinfo
  {author} {\bibfnamefont {X.}~\bibnamefont {Zhang}}, \bibinfo {author}
  {\bibfnamefont {T.}~\bibnamefont {Bothwell}}, \bibinfo {author}
  {\bibfnamefont {J.}~\bibnamefont {Schachenmayer}}, \bibinfo {author}
  {\bibfnamefont {T.~L.}\ \bibnamefont {Nicholson}}, \bibinfo {author}
  {\bibfnamefont {R.}~\bibnamefont {Kaiser}}, \bibinfo {author} {\bibfnamefont
  {S.~F.}\ \bibnamefont {Yelin}}, \bibinfo {author} {\bibfnamefont {M.~D.}\
  \bibnamefont {Lukin}}, \bibinfo {author} {\bibfnamefont {A.~M.}\ \bibnamefont
  {Rey}}, \ and\ \bibinfo {author} {\bibfnamefont {J.}~\bibnamefont {Ye}},\
  }\bibfield  {title} {\enquote {\bibinfo {title} {{Collective atomic
  scattering and motional effects in a dense coherent medium}},}\ }\href
  {\doibase 10.1038/ncomms11039} {\bibfield  {journal} {\bibinfo  {journal}
  {Nat. Comm.}\ }\textbf {\bibinfo {volume} {7}},\ \bibinfo {pages} {11039}
  (\bibinfo {year} {2016})},\ \Eprint {http://arxiv.org/abs/1601.05322}
  {arXiv:1601.05322} \BibitemShut {NoStop}%
\bibitem [{\citenamefont {Peyrot}\ \emph {et~al.}(2018)\citenamefont {Peyrot},
  \citenamefont {Sortais}, \citenamefont {Browaeys}, \citenamefont {Sargsyan},
  \citenamefont {Sarkisyan}, \citenamefont {Keaveney}, \citenamefont {Hughes},\
  and\ \citenamefont {Adams}}]{Peyrot2018}%
  \BibitemOpen
  \bibfield  {author} {\bibinfo {author} {\bibfnamefont {T.}~\bibnamefont
  {Peyrot}}, \bibinfo {author} {\bibfnamefont {Y.~R.~P.}\ \bibnamefont
  {Sortais}}, \bibinfo {author} {\bibfnamefont {A.}~\bibnamefont {Browaeys}},
  \bibinfo {author} {\bibfnamefont {A.}~\bibnamefont {Sargsyan}}, \bibinfo
  {author} {\bibfnamefont {D.}~\bibnamefont {Sarkisyan}}, \bibinfo {author}
  {\bibfnamefont {J.}~\bibnamefont {Keaveney}}, \bibinfo {author}
  {\bibfnamefont {I.~G.}\ \bibnamefont {Hughes}}, \ and\ \bibinfo {author}
  {\bibfnamefont {C.~S.}\ \bibnamefont {Adams}},\ }\bibfield  {title} {\enquote
  {\bibinfo {title} {Collective lamb shift of a nanoscale atomic vapor layer
  within a sapphire cavity},}\ }\href {\doibase 10.1103/PhysRevLett.120.243401}
  {\bibfield  {journal} {\bibinfo  {journal} {Phys. Rev. Lett.}\ }\textbf
  {\bibinfo {volume} {120}},\ \bibinfo {pages} {243401} (\bibinfo {year}
  {2018})}\BibitemShut {NoStop}%
\bibitem [{\citenamefont {Bienaim{\'{e}}}\ \emph {et~al.}(2012)\citenamefont
  {Bienaim{\'{e}}}, \citenamefont {Piovella},\ and\ \citenamefont
  {Kaiser}}]{Bienaime2012}%
  \BibitemOpen
  \bibfield  {author} {\bibinfo {author} {\bibfnamefont {Tom}\ \bibnamefont
  {Bienaim{\'{e}}}}, \bibinfo {author} {\bibfnamefont {Nicola}\ \bibnamefont
  {Piovella}}, \ and\ \bibinfo {author} {\bibfnamefont {Robin}\ \bibnamefont
  {Kaiser}},\ }\bibfield  {title} {\enquote {\bibinfo {title} {{Controlled
  Dicke subradiance from a large cloud of two-level systems}},}\ }\href
  {\doibase 10.1103/PhysRevLett.108.123602} {\bibfield  {journal} {\bibinfo
  {journal} {Phys. Rev. Lett.}\ }\textbf {\bibinfo {volume} {108}},\ \bibinfo
  {pages} {123602} (\bibinfo {year} {2012})},\ \Eprint
  {http://arxiv.org/abs/1201.2274} {arXiv:1201.2274} \BibitemShut {NoStop}%
\bibitem [{\citenamefont {Ott}\ \emph {et~al.}(2013)\citenamefont {Ott},
  \citenamefont {Wubs}, \citenamefont {Lodahl}, \citenamefont {Mortensen},\
  and\ \citenamefont {Kaiser}}]{Ott2013}%
  \BibitemOpen
  \bibfield  {author} {\bibinfo {author} {\bibfnamefont {J.~R.}\ \bibnamefont
  {Ott}}, \bibinfo {author} {\bibfnamefont {M.}~\bibnamefont {Wubs}}, \bibinfo
  {author} {\bibfnamefont {P.}~\bibnamefont {Lodahl}}, \bibinfo {author}
  {\bibfnamefont {N.~A.}\ \bibnamefont {Mortensen}}, \ and\ \bibinfo {author}
  {\bibfnamefont {R.}~\bibnamefont {Kaiser}},\ }\bibfield  {title} {\enquote
  {\bibinfo {title} {Cooperative fluorescence from a strongly driven dilute
  cloud of atoms},}\ }\href {\doibase 10.1103/PhysRevA.87.061801} {\bibfield
  {journal} {\bibinfo  {journal} {Phys. Rev. A}\ }\textbf {\bibinfo {volume}
  {87}},\ \bibinfo {pages} {061801} (\bibinfo {year} {2013})}\BibitemShut
  {NoStop}%
\bibitem [{\citenamefont {de~Oliveira}\ \emph {et~al.}(2014)\citenamefont
  {de~Oliveira}, \citenamefont {Mendes}, \citenamefont {Martins}, \citenamefont
  {Saldanha}, \citenamefont {Tabosa},\ and\ \citenamefont
  {Felinto}}]{Oliveira2014}%
  \BibitemOpen
  \bibfield  {author} {\bibinfo {author} {\bibfnamefont {Rafael~A.}\
  \bibnamefont {de~Oliveira}}, \bibinfo {author} {\bibfnamefont {Milrian~S.}\
  \bibnamefont {Mendes}}, \bibinfo {author} {\bibfnamefont {Weliton~S.}\
  \bibnamefont {Martins}}, \bibinfo {author} {\bibfnamefont {Pablo~L.}\
  \bibnamefont {Saldanha}}, \bibinfo {author} {\bibfnamefont {Jos\'e W.~R.}\
  \bibnamefont {Tabosa}}, \ and\ \bibinfo {author} {\bibfnamefont {Daniel}\
  \bibnamefont {Felinto}},\ }\bibfield  {title} {\enquote {\bibinfo {title}
  {Single-photon superradiance in cold atoms},}\ }\href {\doibase
  10.1103/PhysRevA.90.023848} {\bibfield  {journal} {\bibinfo  {journal} {Phys.
  Rev. A}\ }\textbf {\bibinfo {volume} {90}},\ \bibinfo {pages} {023848}
  (\bibinfo {year} {2014})}\BibitemShut {NoStop}%
\bibitem [{\citenamefont {Guerin}\ \emph {et~al.}(2016)\citenamefont {Guerin},
  \citenamefont {Ara{\'{u}}jo},\ and\ \citenamefont {Kaiser}}]{Guerin2016}%
  \BibitemOpen
  \bibfield  {author} {\bibinfo {author} {\bibfnamefont {William}\ \bibnamefont
  {Guerin}}, \bibinfo {author} {\bibfnamefont {Michelle~O.}\ \bibnamefont
  {Ara{\'{u}}jo}}, \ and\ \bibinfo {author} {\bibfnamefont {Robin}\
  \bibnamefont {Kaiser}},\ }\bibfield  {title} {\enquote {\bibinfo {title}
  {{Subradiance in a Large Cloud of Cold Atoms}},}\ }\href {\doibase
  10.1103/PhysRevLett.116.083601} {\bibfield  {journal} {\bibinfo  {journal}
  {Phys. Rev. Lett.}\ }\textbf {\bibinfo {volume} {116}},\ \bibinfo {pages}
  {083601} (\bibinfo {year} {2016})},\ \Eprint
  {http://arxiv.org/abs/1509.00227} {arXiv:1509.00227} \BibitemShut {NoStop}%
\bibitem [{\citenamefont {Ara\'ujo}\ \emph {et~al.}(2016)\citenamefont
  {Ara\'ujo}, \citenamefont {Kre\ifmmode \check{s}\else
  \v{s}\fi{}i\ifmmode~\acute{c}\else \'{c}\fi{}}, \citenamefont {Kaiser},\ and\
  \citenamefont {Guerin}}]{Araujo2016}%
  \BibitemOpen
  \bibfield  {author} {\bibinfo {author} {\bibfnamefont {Michelle~O.}\
  \bibnamefont {Ara\'ujo}}, \bibinfo {author} {\bibfnamefont {Ivor}\
  \bibnamefont {Kre\ifmmode \check{s}\else \v{s}\fi{}i\ifmmode~\acute{c}\else
  \'{c}\fi{}}}, \bibinfo {author} {\bibfnamefont {Robin}\ \bibnamefont
  {Kaiser}}, \ and\ \bibinfo {author} {\bibfnamefont {William}\ \bibnamefont
  {Guerin}},\ }\bibfield  {title} {\enquote {\bibinfo {title} {Superradiance in
  a large and dilute cloud of cold atoms in the linear-optics regime},}\ }\href
  {\doibase 10.1103/PhysRevLett.117.073002} {\bibfield  {journal} {\bibinfo
  {journal} {Phys. Rev. Lett.}\ }\textbf {\bibinfo {volume} {117}},\ \bibinfo
  {pages} {073002} (\bibinfo {year} {2016})}\BibitemShut {NoStop}%
\bibitem [{\citenamefont {Roof}\ \emph {et~al.}(2016)\citenamefont {Roof},
  \citenamefont {Kemp}, \citenamefont {Havey},\ and\ \citenamefont
  {Sokolov}}]{Roof2016}%
  \BibitemOpen
  \bibfield  {author} {\bibinfo {author} {\bibfnamefont {S.~J.}\ \bibnamefont
  {Roof}}, \bibinfo {author} {\bibfnamefont {K.~J.}\ \bibnamefont {Kemp}},
  \bibinfo {author} {\bibfnamefont {M.~D.}\ \bibnamefont {Havey}}, \ and\
  \bibinfo {author} {\bibfnamefont {I.~M.}\ \bibnamefont {Sokolov}},\
  }\bibfield  {title} {\enquote {\bibinfo {title} {Observation of single-photon
  superradiance and the cooperative lamb shift in an extended sample of cold
  atoms},}\ }\href {\doibase 10.1103/PhysRevLett.117.073003} {\bibfield
  {journal} {\bibinfo  {journal} {Phys. Rev. Lett.}\ }\textbf {\bibinfo
  {volume} {117}},\ \bibinfo {pages} {073003} (\bibinfo {year}
  {2016})}\BibitemShut {NoStop}%
\bibitem [{\citenamefont {Ara\'{u}jo}\ \emph {et~al.}(2018)\citenamefont
  {Ara\'{u}jo}, \citenamefont {Guerin},\ and\ \citenamefont
  {Kaiser}}]{Araujo2017}%
  \BibitemOpen
  \bibfield  {author} {\bibinfo {author} {\bibfnamefont {M.~O.}\ \bibnamefont
  {Ara\'{u}jo}}, \bibinfo {author} {\bibfnamefont {W.}~\bibnamefont {Guerin}},
  \ and\ \bibinfo {author} {\bibfnamefont {R.}~\bibnamefont {Kaiser}},\
  }\bibfield  {title} {\enquote {\bibinfo {title} {Decay dynamics in the
  coupled-dipole model},}\ }\href {\doibase 10.1080/09500340.2017.1380856}
  {\bibfield  {journal} {\bibinfo  {journal} {J. Mod. Opt.}\ }\textbf {\bibinfo
  {volume} {65}},\ \bibinfo {pages} {1345} (\bibinfo {year} {2018})},\ \Eprint
  {http://arxiv.org/abs/https://doi.org/10.1080/09500340.2017.1380856}
  {https://doi.org/10.1080/09500340.2017.1380856} \BibitemShut {NoStop}%
\bibitem [{\citenamefont {Cottier}\ \emph {et~al.}(2018)\citenamefont
  {Cottier}, \citenamefont {Kaiser},\ and\ \citenamefont
  {Bachelard}}]{Cottier2018}%
  \BibitemOpen
  \bibfield  {author} {\bibinfo {author} {\bibfnamefont {Florent}\ \bibnamefont
  {Cottier}}, \bibinfo {author} {\bibfnamefont {Robin}\ \bibnamefont {Kaiser}},
  \ and\ \bibinfo {author} {\bibfnamefont {Romain}\ \bibnamefont {Bachelard}},\
  }\bibfield  {title} {\enquote {\bibinfo {title} {Role of disorder in super-
  and subradiance of cold atomic clouds},}\ }\href {\doibase
  10.1103/PhysRevA.98.013622} {\bibfield  {journal} {\bibinfo  {journal} {Phys.
  Rev. A}\ }\textbf {\bibinfo {volume} {98}},\ \bibinfo {pages} {013622}
  (\bibinfo {year} {2018})}\BibitemShut {NoStop}%
\bibitem [{\citenamefont {Filipp}\ \emph {et~al.}(2011)\citenamefont {Filipp},
  \citenamefont {van Loo}, \citenamefont {Baur}, \citenamefont {Steffen},\ and\
  \citenamefont {Wallraff}}]{Filipp2011}%
  \BibitemOpen
  \bibfield  {author} {\bibinfo {author} {\bibfnamefont {S.}~\bibnamefont
  {Filipp}}, \bibinfo {author} {\bibfnamefont {A.~F.}\ \bibnamefont {van Loo}},
  \bibinfo {author} {\bibfnamefont {M.}~\bibnamefont {Baur}}, \bibinfo {author}
  {\bibfnamefont {L.}~\bibnamefont {Steffen}}, \ and\ \bibinfo {author}
  {\bibfnamefont {A.}~\bibnamefont {Wallraff}},\ }\bibfield  {title} {\enquote
  {\bibinfo {title} {Preparation of subradiant states using local qubit control
  in circuit qed},}\ }\href {\doibase 10.1103/PhysRevA.84.061805} {\bibfield
  {journal} {\bibinfo  {journal} {Phys. Rev. A}\ }\textbf {\bibinfo {volume}
  {84}},\ \bibinfo {pages} {061805} (\bibinfo {year} {2011})}\BibitemShut
  {NoStop}%
\bibitem [{\citenamefont {Lalumi\`ere}\ \emph {et~al.}(2013)\citenamefont
  {Lalumi\`ere}, \citenamefont {Sanders}, \citenamefont {van Loo},
  \citenamefont {Fedorov}, \citenamefont {Wallraff},\ and\ \citenamefont
  {Blais}}]{Lalumiere2013}%
  \BibitemOpen
  \bibfield  {author} {\bibinfo {author} {\bibfnamefont {Kevin}\ \bibnamefont
  {Lalumi\`ere}}, \bibinfo {author} {\bibfnamefont {Barry~C.}\ \bibnamefont
  {Sanders}}, \bibinfo {author} {\bibfnamefont {A.~F.}\ \bibnamefont {van
  Loo}}, \bibinfo {author} {\bibfnamefont {A.}~\bibnamefont {Fedorov}},
  \bibinfo {author} {\bibfnamefont {A.}~\bibnamefont {Wallraff}}, \ and\
  \bibinfo {author} {\bibfnamefont {A.}~\bibnamefont {Blais}},\ }\bibfield
  {title} {\enquote {\bibinfo {title} {Input-output theory for waveguide qed
  with an ensemble of inhomogeneous atoms},}\ }\href {\doibase
  10.1103/PhysRevA.88.043806} {\bibfield  {journal} {\bibinfo  {journal} {Phys.
  Rev. A}\ }\textbf {\bibinfo {volume} {88}},\ \bibinfo {pages} {043806}
  (\bibinfo {year} {2013})}\BibitemShut {NoStop}%
\bibitem [{\citenamefont {Jenkins}\ \emph {et~al.}(2017)\citenamefont
  {Jenkins}, \citenamefont {Ruostekoski}, \citenamefont {Papasimakis},
  \citenamefont {Savo},\ and\ \citenamefont {Zheludev}}]{Jenkins2017}%
  \BibitemOpen
  \bibfield  {author} {\bibinfo {author} {\bibfnamefont {Stewart~D.}\
  \bibnamefont {Jenkins}}, \bibinfo {author} {\bibfnamefont {Janne}\
  \bibnamefont {Ruostekoski}}, \bibinfo {author} {\bibfnamefont {Nikitas}\
  \bibnamefont {Papasimakis}}, \bibinfo {author} {\bibfnamefont {Salvatore}\
  \bibnamefont {Savo}}, \ and\ \bibinfo {author} {\bibfnamefont {Nikolay~I.}\
  \bibnamefont {Zheludev}},\ }\bibfield  {title} {\enquote {\bibinfo {title}
  {Many-body subradiant excitations in metamaterial arrays: Experiment and
  theory},}\ }\href {\doibase 10.1103/PhysRevLett.119.053901} {\bibfield
  {journal} {\bibinfo  {journal} {Phys. Rev. Lett.}\ }\textbf {\bibinfo
  {volume} {119}},\ \bibinfo {pages} {053901} (\bibinfo {year}
  {2017})}\BibitemShut {NoStop}%
\bibitem [{\citenamefont {Lodahl}\ \emph {et~al.}(2004)\citenamefont {Lodahl},
  \citenamefont {Floris~van Driel}, \citenamefont {Nikolaev}, \citenamefont
  {Irman}, \citenamefont {Overgaag}, \citenamefont {Vanmaekelbergh},\ and\
  \citenamefont {Vos}}]{Lodahl2004}%
  \BibitemOpen
  \bibfield  {author} {\bibinfo {author} {\bibfnamefont {Peter}\ \bibnamefont
  {Lodahl}}, \bibinfo {author} {\bibfnamefont {A.}~\bibnamefont {Floris~van
  Driel}}, \bibinfo {author} {\bibfnamefont {Ivan~S.}\ \bibnamefont
  {Nikolaev}}, \bibinfo {author} {\bibfnamefont {Arie}\ \bibnamefont {Irman}},
  \bibinfo {author} {\bibfnamefont {Karin}\ \bibnamefont {Overgaag}}, \bibinfo
  {author} {\bibfnamefont {Dani\"{e}l}\ \bibnamefont {Vanmaekelbergh}}, \ and\
  \bibinfo {author} {\bibfnamefont {Willem~L.}\ \bibnamefont {Vos}},\
  }\bibfield  {title} {\enquote {\bibinfo {title} {Controlling the dynamics of
  spontaneous emission from quantum dots by photonic crystals},}\ }\href
  {\doibase 10.1038/nature02772} {\bibfield  {journal} {\bibinfo  {journal}
  {Nature}\ }\textbf {\bibinfo {volume} {430}},\ \bibinfo {pages} {654}
  (\bibinfo {year} {2004})}\BibitemShut {NoStop}%
\bibitem [{\citenamefont {Tighineanu}\ \emph {et~al.}(2016)\citenamefont
  {Tighineanu}, \citenamefont {Daveau}, \citenamefont {Lehmann}, \citenamefont
  {Beere}, \citenamefont {Ritchie}, \citenamefont {Lodahl},\ and\ \citenamefont
  {Stobbe}}]{Tighineanu2016}%
  \BibitemOpen
  \bibfield  {author} {\bibinfo {author} {\bibfnamefont {Petru}\ \bibnamefont
  {Tighineanu}}, \bibinfo {author} {\bibfnamefont {Rapha\"el~S.}\ \bibnamefont
  {Daveau}}, \bibinfo {author} {\bibfnamefont {Tau~B.}\ \bibnamefont
  {Lehmann}}, \bibinfo {author} {\bibfnamefont {Harvey~E.}\ \bibnamefont
  {Beere}}, \bibinfo {author} {\bibfnamefont {David~A.}\ \bibnamefont
  {Ritchie}}, \bibinfo {author} {\bibfnamefont {Peter}\ \bibnamefont {Lodahl}},
  \ and\ \bibinfo {author} {\bibfnamefont {S\o{}ren}\ \bibnamefont {Stobbe}},\
  }\bibfield  {title} {\enquote {\bibinfo {title} {Single-photon superradiance
  from a quantum dot},}\ }\href {\doibase 10.1103/PhysRevLett.116.163604}
  {\bibfield  {journal} {\bibinfo  {journal} {Phys. Rev. Lett.}\ }\textbf
  {\bibinfo {volume} {116}},\ \bibinfo {pages} {163604} (\bibinfo {year}
  {2016})}\BibitemShut {NoStop}%
\bibitem [{\citenamefont {Kim}\ \emph {et~al.}(2018)\citenamefont {Kim},
  \citenamefont {Aghaeimeibodi}, \citenamefont {Richardson}, \citenamefont
  {Leavitt},\ and\ \citenamefont {Waks}}]{Kim2018}%
  \BibitemOpen
  \bibfield  {author} {\bibinfo {author} {\bibfnamefont {Je-Hyung}\
  \bibnamefont {Kim}}, \bibinfo {author} {\bibfnamefont {Shahriar}\
  \bibnamefont {Aghaeimeibodi}}, \bibinfo {author} {\bibfnamefont {Christopher
  J.~K.}\ \bibnamefont {Richardson}}, \bibinfo {author} {\bibfnamefont
  {Richard~P.}\ \bibnamefont {Leavitt}}, \ and\ \bibinfo {author}
  {\bibfnamefont {Edo}\ \bibnamefont {Waks}},\ }\bibfield  {title} {\enquote
  {\bibinfo {title} {Super-radiant emission from quantum dots in a nanophotonic
  waveguide},}\ }\href {\doibase 10.1021/acs.nanolett.8b01133} {\bibfield
  {journal} {\bibinfo  {journal} {Nano Lett.}\ }\textbf {\bibinfo {volume}
  {18}},\ \bibinfo {pages} {4734} (\bibinfo {year} {2018})},\ \bibinfo {note}
  {pMID: 29966093},\ \Eprint
  {http://arxiv.org/abs/https://doi.org/10.1021/acs.nanolett.8b01133}
  {https://doi.org/10.1021/acs.nanolett.8b01133} \BibitemShut {NoStop}%
\bibitem [{\citenamefont {Scully}\ \emph {et~al.}(2006)\citenamefont {Scully},
  \citenamefont {Fry}, \citenamefont {Ooi},\ and\ \citenamefont
  {W\'odkiewicz}}]{Scully2006}%
  \BibitemOpen
  \bibfield  {author} {\bibinfo {author} {\bibfnamefont {Marlan~O.}\
  \bibnamefont {Scully}}, \bibinfo {author} {\bibfnamefont {Edward~S.}\
  \bibnamefont {Fry}}, \bibinfo {author} {\bibfnamefont {C.~H.~Raymond}\
  \bibnamefont {Ooi}}, \ and\ \bibinfo {author} {\bibfnamefont {Krzysztof}\
  \bibnamefont {W\'odkiewicz}},\ }\bibfield  {title} {\enquote {\bibinfo
  {title} {Directed spontaneous emission from an extended ensemble of $n$
  atoms: Timing is everything},}\ }\href {\doibase
  10.1103/PhysRevLett.96.010501} {\bibfield  {journal} {\bibinfo  {journal}
  {Phys. Rev. Lett.}\ }\textbf {\bibinfo {volume} {96}},\ \bibinfo {pages}
  {010501} (\bibinfo {year} {2006})}\BibitemShut {NoStop}%
\bibitem [{\citenamefont {Plankensteiner}\ \emph {et~al.}(2015)\citenamefont
  {Plankensteiner}, \citenamefont {Ostermann}, \citenamefont {Ritsch},\ and\
  \citenamefont {Genes}}]{Plankensteiner2015}%
  \BibitemOpen
  \bibfield  {author} {\bibinfo {author} {\bibfnamefont {D.}~\bibnamefont
  {Plankensteiner}}, \bibinfo {author} {\bibfnamefont {L.}~\bibnamefont
  {Ostermann}}, \bibinfo {author} {\bibfnamefont {H.}~\bibnamefont {Ritsch}}, \
  and\ \bibinfo {author} {\bibfnamefont {C.}~\bibnamefont {Genes}},\ }\bibfield
   {title} {\enquote {\bibinfo {title} {{Selective protected state preparation
  of coupled dissipative quantum emitters}},}\ }\href {\doibase
  10.1038/srep16231} {\bibfield  {journal} {\bibinfo  {journal} {Sci. Rep.}\
  }\textbf {\bibinfo {volume} {5}},\ \bibinfo {pages} {16231} (\bibinfo {year}
  {2015})},\ \Eprint {http://arxiv.org/abs/1504.08148} {arXiv:1504.08148}
  \BibitemShut {NoStop}%
\bibitem [{\citenamefont {Scully}(2015)}]{Scully2015}%
  \BibitemOpen
  \bibfield  {author} {\bibinfo {author} {\bibfnamefont {Marlan~O.}\
  \bibnamefont {Scully}},\ }\bibfield  {title} {\enquote {\bibinfo {title}
  {Single photon subradiance: Quantum control of spontaneous emission and
  ultrafast readout},}\ }\href {\doibase 10.1103/PhysRevLett.115.243602}
  {\bibfield  {journal} {\bibinfo  {journal} {Phys. Rev. Lett.}\ }\textbf
  {\bibinfo {volume} {115}},\ \bibinfo {pages} {243602} (\bibinfo {year}
  {2015})}\BibitemShut {NoStop}%
\bibitem [{\citenamefont {Jen}\ \emph {et~al.}(2016)\citenamefont {Jen},
  \citenamefont {Chang},\ and\ \citenamefont {Chen}}]{Jen2016}%
  \BibitemOpen
  \bibfield  {author} {\bibinfo {author} {\bibfnamefont {H.~H.}\ \bibnamefont
  {Jen}}, \bibinfo {author} {\bibfnamefont {M.~S.}\ \bibnamefont {Chang}}, \
  and\ \bibinfo {author} {\bibfnamefont {Y.~C.}\ \bibnamefont {Chen}},\
  }\bibfield  {title} {\enquote {\bibinfo {title} {{Cooperative single-photon
  subradiant states}},}\ }\href {\doibase 10.1103/PhysRevA.94.013803}
  {\bibfield  {journal} {\bibinfo  {journal} {Phys. Rev. A}\ }\textbf {\bibinfo
  {volume} {94}},\ \bibinfo {pages} {013803} (\bibinfo {year} {2016})},\
  \Eprint {http://arxiv.org/abs/1603.00996} {arXiv:1603.00996} \BibitemShut
  {NoStop}%
\bibitem [{\citenamefont {Facchinetti}\ \emph {et~al.}(2016)\citenamefont
  {Facchinetti}, \citenamefont {Jenkins},\ and\ \citenamefont
  {Ruostekoski}}]{Facchinetti2016}%
  \BibitemOpen
  \bibfield  {author} {\bibinfo {author} {\bibfnamefont {G.}~\bibnamefont
  {Facchinetti}}, \bibinfo {author} {\bibfnamefont {S.~D.}\ \bibnamefont
  {Jenkins}}, \ and\ \bibinfo {author} {\bibfnamefont {J.}~\bibnamefont
  {Ruostekoski}},\ }\bibfield  {title} {\enquote {\bibinfo {title} {{Storing
  Light with Subradiant Correlations in Arrays of Atoms}},}\ }\href {\doibase
  10.1103/PhysRevLett.117.243601} {\bibfield  {journal} {\bibinfo  {journal}
  {Phys. Rev. Lett.}\ }\textbf {\bibinfo {volume} {117}},\ \bibinfo {pages}
  {243601} (\bibinfo {year} {2016})},\ \Eprint
  {http://arxiv.org/abs/1609.08350} {arXiv:1609.08350} \BibitemShut {NoStop}%
\bibitem [{\citenamefont {Asenjo-Garcia}\ \emph {et~al.}(2017)\citenamefont
  {Asenjo-Garcia}, \citenamefont {Moreno-Cardoner}, \citenamefont {Albrecht},
  \citenamefont {Kimble},\ and\ \citenamefont {Chang}}]{Asenjo-Garcia2017}%
  \BibitemOpen
  \bibfield  {author} {\bibinfo {author} {\bibfnamefont {A.}~\bibnamefont
  {Asenjo-Garcia}}, \bibinfo {author} {\bibfnamefont {M.}~\bibnamefont
  {Moreno-Cardoner}}, \bibinfo {author} {\bibfnamefont {A.}~\bibnamefont
  {Albrecht}}, \bibinfo {author} {\bibfnamefont {H.~J.}\ \bibnamefont
  {Kimble}}, \ and\ \bibinfo {author} {\bibfnamefont {D.~E.}\ \bibnamefont
  {Chang}},\ }\bibfield  {title} {\enquote {\bibinfo {title} {{Exponential
  improvement in photon storage fidelities using subradiance and "selective
  radiance" in atomic arrays}},}\ }\href {\doibase 10.1103/PhysRevX.7.031024}
  {\bibfield  {journal} {\bibinfo  {journal} {Phys. Rev. X}\ }\textbf {\bibinfo
  {volume} {7}},\ \bibinfo {pages} {031024} (\bibinfo {year} {2017})},\ \Eprint
  {http://arxiv.org/abs/1703.03382} {arXiv:1703.03382} \BibitemShut {NoStop}%
\bibitem [{\citenamefont {Jen}(2017)}]{Jen2017}%
  \BibitemOpen
  \bibfield  {author} {\bibinfo {author} {\bibfnamefont {H.~H.}\ \bibnamefont
  {Jen}},\ }\bibfield  {title} {\enquote {\bibinfo {title} {{Phase-imprinted
  multiphoton subradiant states}},}\ }\href {\doibase
  10.1103/PhysRevA.96.023814} {\bibfield  {journal} {\bibinfo  {journal} {Phys.
  Rev. A}\ }\textbf {\bibinfo {volume} {96}},\ \bibinfo {pages} {023814}
  (\bibinfo {year} {2017})},\ \Eprint {http://arxiv.org/abs/1706.00888}
  {arXiv:1706.00888} \BibitemShut {NoStop}%
\bibitem [{\citenamefont {Jen}(2018)}]{Jen2018}%
  \BibitemOpen
  \bibfield  {author} {\bibinfo {author} {\bibfnamefont {H.~H.}\ \bibnamefont
  {Jen}},\ }\bibfield  {title} {\enquote {\bibinfo {title} {{Directional
  subradiance from helical-phase-imprinted multiphoton states}},}\ }\href
  {\doibase 10.1038/s41598-018-25592-5} {\bibfield  {journal} {\bibinfo
  {journal} {Sci. Rep.}\ }\textbf {\bibinfo {volume} {8}},\ \bibinfo {pages}
  {7163} (\bibinfo {year} {2018})},\ \Eprint {http://arxiv.org/abs/1801.02834}
  {arXiv:1801.02834} \BibitemShut {NoStop}%
\bibitem [{\citenamefont {Jen}\ \emph {et~al.}(2018)\citenamefont {Jen},
  \citenamefont {Chang},\ and\ \citenamefont {Chen}}]{Jen2018-1}%
  \BibitemOpen
  \bibfield  {author} {\bibinfo {author} {\bibfnamefont {H.~H.}\ \bibnamefont
  {Jen}}, \bibinfo {author} {\bibfnamefont {M.-S.}\ \bibnamefont {Chang}}, \
  and\ \bibinfo {author} {\bibfnamefont {Y.-C.}\ \bibnamefont {Chen}},\
  }\bibfield  {title} {\enquote {\bibinfo {title} {{Cooperative light
  scattering from helical-phase-imprinted atomic rings}},}\ }\href {\doibase
  10.1038/s41598-018-27888-y} {\bibfield  {journal} {\bibinfo  {journal} {Sci.
  Rep.}\ }\textbf {\bibinfo {volume} {8}},\ \bibinfo {pages} {9570} (\bibinfo
  {year} {2018})}\BibitemShut {NoStop}%
\bibitem [{\citenamefont {Moreno-Cardoner}\ \emph {et~al.}(2019)\citenamefont
  {Moreno-Cardoner}, \citenamefont {Plankensteiner}, \citenamefont {Ostermann},
  \citenamefont {Chang},\ and\ \citenamefont {Ritsch}}]{Moreno2019}%
  \BibitemOpen
  \bibfield  {author} {\bibinfo {author} {\bibfnamefont {Maria}\ \bibnamefont
  {Moreno-Cardoner}}, \bibinfo {author} {\bibfnamefont {David}\ \bibnamefont
  {Plankensteiner}}, \bibinfo {author} {\bibfnamefont {Laurin}\ \bibnamefont
  {Ostermann}}, \bibinfo {author} {\bibfnamefont {Darrick}\ \bibnamefont
  {Chang}}, \ and\ \bibinfo {author} {\bibfnamefont {Helmut}\ \bibnamefont
  {Ritsch}},\ }\bibfield  {title} {\enquote {\bibinfo {title} {{Extraordinary
  subradiance with lossless excitation transfer in dipole-coupled nano-rings of
  quantum emitters}},}\ }\href@noop {} {\bibfield  {journal} {\bibinfo
  {journal} {arXiv:1901.10598}\ } (\bibinfo {year} {2019})}\BibitemShut
  {NoStop}%
\bibitem [{\citenamefont {Guimond}\ \emph {et~al.}(2019)\citenamefont
  {Guimond}, \citenamefont {Grankin}, \citenamefont {Vasilyev}, \citenamefont
  {Vermersch},\ and\ \citenamefont {Zoller}}]{Guimond2019}%
  \BibitemOpen
  \bibfield  {author} {\bibinfo {author} {\bibfnamefont {P.-O.}\ \bibnamefont
  {Guimond}}, \bibinfo {author} {\bibfnamefont {A.}~\bibnamefont {Grankin}},
  \bibinfo {author} {\bibfnamefont {D.~V.}\ \bibnamefont {Vasilyev}}, \bibinfo
  {author} {\bibfnamefont {B.}~\bibnamefont {Vermersch}}, \ and\ \bibinfo
  {author} {\bibfnamefont {P.}~\bibnamefont {Zoller}},\ }\bibfield  {title}
  {\enquote {\bibinfo {title} {Subradiant bell states in distant atomic
  arrays},}\ }\href {\doibase 10.1103/PhysRevLett.122.093601} {\bibfield
  {journal} {\bibinfo  {journal} {Phys. Rev. Lett.}\ }\textbf {\bibinfo
  {volume} {122}},\ \bibinfo {pages} {093601} (\bibinfo {year}
  {2019})}\BibitemShut {NoStop}%
\bibitem [{\citenamefont {Bettles}\ \emph {et~al.}(2017)\citenamefont
  {Bettles}, \citenamefont {Min\'a\ifmmode~\check{r}\else \v{r}\fi{}},
  \citenamefont {Adams}, \citenamefont {Lesanovsky},\ and\ \citenamefont
  {Olmos}}]{Bettles2017}%
  \BibitemOpen
  \bibfield  {author} {\bibinfo {author} {\bibfnamefont {Robert~J.}\
  \bibnamefont {Bettles}}, \bibinfo {author} {\bibfnamefont {Ji\ifmmode
  \check{r}\else~\v{r}\fi{}\'{\i}}\ \bibnamefont {Min\'a\ifmmode~\check{r}\else
  \v{r}\fi{}}}, \bibinfo {author} {\bibfnamefont {Charles~S.}\ \bibnamefont
  {Adams}}, \bibinfo {author} {\bibfnamefont {Igor}\ \bibnamefont
  {Lesanovsky}}, \ and\ \bibinfo {author} {\bibfnamefont {Beatriz}\
  \bibnamefont {Olmos}},\ }\bibfield  {title} {\enquote {\bibinfo {title}
  {Topological properties of a dense atomic lattice gas},}\ }\href {\doibase
  10.1103/PhysRevA.96.041603} {\bibfield  {journal} {\bibinfo  {journal} {Phys.
  Rev. A}\ }\textbf {\bibinfo {volume} {96}},\ \bibinfo {pages} {041603}
  (\bibinfo {year} {2017})}\BibitemShut {NoStop}%
\bibitem [{\citenamefont {Perczel}\ \emph {et~al.}(2017)\citenamefont
  {Perczel}, \citenamefont {Borregaard}, \citenamefont {Chang}, \citenamefont
  {Pichler}, \citenamefont {Yelin}, \citenamefont {Zoller},\ and\ \citenamefont
  {Lukin}}]{Perczel2017}%
  \BibitemOpen
  \bibfield  {author} {\bibinfo {author} {\bibfnamefont {J.}~\bibnamefont
  {Perczel}}, \bibinfo {author} {\bibfnamefont {J.}~\bibnamefont {Borregaard}},
  \bibinfo {author} {\bibfnamefont {D.~E.}\ \bibnamefont {Chang}}, \bibinfo
  {author} {\bibfnamefont {H.}~\bibnamefont {Pichler}}, \bibinfo {author}
  {\bibfnamefont {S.~F.}\ \bibnamefont {Yelin}}, \bibinfo {author}
  {\bibfnamefont {P.}~\bibnamefont {Zoller}}, \ and\ \bibinfo {author}
  {\bibfnamefont {M.~D.}\ \bibnamefont {Lukin}},\ }\bibfield  {title} {\enquote
  {\bibinfo {title} {Topological quantum optics in two-dimensional atomic
  arrays},}\ }\href {\doibase 10.1103/PhysRevLett.119.023603} {\bibfield
  {journal} {\bibinfo  {journal} {Phys. Rev. Lett.}\ }\textbf {\bibinfo
  {volume} {119}},\ \bibinfo {pages} {023603} (\bibinfo {year}
  {2017})}\BibitemShut {NoStop}%
\bibitem [{\citenamefont {Willingham}\ and\ \citenamefont
  {Link}(2011)}]{Willingham2011}%
  \BibitemOpen
  \bibfield  {author} {\bibinfo {author} {\bibfnamefont {Britain}\ \bibnamefont
  {Willingham}}\ and\ \bibinfo {author} {\bibfnamefont {Stephan}\ \bibnamefont
  {Link}},\ }\bibfield  {title} {\enquote {\bibinfo {title} {Energy transport
  in metal nanoparticle chains via sub-radiant plasmon modes},}\ }\href
  {\doibase 10.1364/OE.19.006450} {\bibfield  {journal} {\bibinfo  {journal}
  {Opt. Express}\ }\textbf {\bibinfo {volume} {19}},\ \bibinfo {pages} {6450}
  (\bibinfo {year} {2011})}\BibitemShut {NoStop}%
\bibitem [{\citenamefont {Giusteri}\ \emph {et~al.}(2015)\citenamefont
  {Giusteri}, \citenamefont {Mattiotti},\ and\ \citenamefont
  {Celardo}}]{Giusteri2015}%
  \BibitemOpen
  \bibfield  {author} {\bibinfo {author} {\bibfnamefont {Giulio~G.}\
  \bibnamefont {Giusteri}}, \bibinfo {author} {\bibfnamefont {Francesco}\
  \bibnamefont {Mattiotti}}, \ and\ \bibinfo {author} {\bibfnamefont {G.~Luca}\
  \bibnamefont {Celardo}},\ }\bibfield  {title} {\enquote {\bibinfo {title}
  {Non-hermitian hamiltonian approach to quantum transport in disordered
  networks with sinks: Validity and effectiveness},}\ }\href {\doibase
  10.1103/PhysRevB.91.094301} {\bibfield  {journal} {\bibinfo  {journal} {Phys.
  Rev. B}\ }\textbf {\bibinfo {volume} {91}},\ \bibinfo {pages} {094301}
  (\bibinfo {year} {2015})}\BibitemShut {NoStop}%
\bibitem [{\citenamefont {Leggio}\ \emph {et~al.}(2015)\citenamefont {Leggio},
  \citenamefont {Messina},\ and\ \citenamefont {Antezza}}]{Leggio2015}%
  \BibitemOpen
  \bibfield  {author} {\bibinfo {author} {\bibfnamefont {B.}~\bibnamefont
  {Leggio}}, \bibinfo {author} {\bibfnamefont {R.}~\bibnamefont {Messina}}, \
  and\ \bibinfo {author} {\bibfnamefont {M.}~\bibnamefont {Antezza}},\
  }\bibfield  {title} {\enquote {\bibinfo {title} {Thermally activated nonlocal
  amplification in quantum energy transport},}\ }\href {\doibase
  10.1209/0295-5075/110/40002} {\bibfield  {journal} {\bibinfo  {journal}
  {Europhys. Lett.}\ }\textbf {\bibinfo {volume} {110}},\ \bibinfo {pages}
  {40002} (\bibinfo {year} {2015})}\BibitemShut {NoStop}%
\bibitem [{\citenamefont {Doyeux}\ \emph {et~al.}(2017)\citenamefont {Doyeux},
  \citenamefont {Messina}, \citenamefont {Leggio},\ and\ \citenamefont
  {Antezza}}]{Doyeux2017}%
  \BibitemOpen
  \bibfield  {author} {\bibinfo {author} {\bibfnamefont {Pierre}\ \bibnamefont
  {Doyeux}}, \bibinfo {author} {\bibfnamefont {Riccardo}\ \bibnamefont
  {Messina}}, \bibinfo {author} {\bibfnamefont {Bruno}\ \bibnamefont {Leggio}},
  \ and\ \bibinfo {author} {\bibfnamefont {Mauro}\ \bibnamefont {Antezza}},\
  }\bibfield  {title} {\enquote {\bibinfo {title} {Excitation injector in an
  atomic chain: Long-range transport and efficiency amplification},}\ }\href
  {\doibase 10.1103/PhysRevA.95.012138} {\bibfield  {journal} {\bibinfo
  {journal} {Phys. Rev. A}\ }\textbf {\bibinfo {volume} {95}},\ \bibinfo
  {pages} {012138} (\bibinfo {year} {2017})}\BibitemShut {NoStop}%
\bibitem [{\citenamefont {Deng}\ \emph {et~al.}(2018)\citenamefont {Deng},
  \citenamefont {Kravtsov}, \citenamefont {Shlyapnikov},\ and\ \citenamefont
  {Santos}}]{Deng2018}%
  \BibitemOpen
  \bibfield  {author} {\bibinfo {author} {\bibfnamefont {X.}~\bibnamefont
  {Deng}}, \bibinfo {author} {\bibfnamefont {V.~E.}\ \bibnamefont {Kravtsov}},
  \bibinfo {author} {\bibfnamefont {G.~V.}\ \bibnamefont {Shlyapnikov}}, \ and\
  \bibinfo {author} {\bibfnamefont {L.}~\bibnamefont {Santos}},\ }\bibfield
  {title} {\enquote {\bibinfo {title} {Duality in power-law localization in
  disordered one-dimensional systems},}\ }\href {\doibase
  10.1103/PhysRevLett.120.110602} {\bibfield  {journal} {\bibinfo  {journal}
  {Phys. Rev. Lett.}\ }\textbf {\bibinfo {volume} {120}},\ \bibinfo {pages}
  {110602} (\bibinfo {year} {2018})}\BibitemShut {NoStop}%
\bibitem [{\citenamefont {Botzung}\ \emph {et~al.}(2018)\citenamefont
  {Botzung}, \citenamefont {Vodola}, \citenamefont {Naldesi}, \citenamefont
  {M\"uller}, \citenamefont {Ercolessi},\ and\ \citenamefont
  {Pupillo}}]{Botzung2018}%
  \BibitemOpen
  \bibfield  {author} {\bibinfo {author} {\bibfnamefont {Thomas}\ \bibnamefont
  {Botzung}}, \bibinfo {author} {\bibfnamefont {Davide}\ \bibnamefont
  {Vodola}}, \bibinfo {author} {\bibfnamefont {Piero}\ \bibnamefont {Naldesi}},
  \bibinfo {author} {\bibfnamefont {Markus}\ \bibnamefont {M\"uller}}, \bibinfo
  {author} {\bibfnamefont {Elisa}\ \bibnamefont {Ercolessi}}, \ and\ \bibinfo
  {author} {\bibfnamefont {Guido}\ \bibnamefont {Pupillo}},\ }\bibfield
  {title} {\enquote {\bibinfo {title} {Algebraic localization from power-law
  interactions in disordered quantum wires},}\ }\href@noop {} {\bibfield
  {journal} {\bibinfo  {journal} {arXiv:1810.09779}\ } (\bibinfo {year}
  {2018})}\BibitemShut {NoStop}%
\bibitem [{\citenamefont {Akkermans}\ and\ \citenamefont
  {Gero}(2013)}]{Akkermans2013}%
  \BibitemOpen
  \bibfield  {author} {\bibinfo {author} {\bibfnamefont {E.}~\bibnamefont
  {Akkermans}}\ and\ \bibinfo {author} {\bibfnamefont {A.}~\bibnamefont
  {Gero}},\ }\bibfield  {title} {\enquote {\bibinfo {title} {Cooperative
  effects in one-dimensional random atomic gases: Absence of single-atom
  limit},}\ }\href {\doibase 10.1209/0295-5075/101/54003} {\bibfield  {journal}
  {\bibinfo  {journal} {Europhys. Lett.}\ }\textbf {\bibinfo {volume} {101}},\
  \bibinfo {pages} {54003} (\bibinfo {year} {2013})}\BibitemShut {NoStop}%
\bibitem [{\citenamefont {Biella}\ \emph {et~al.}(2013)\citenamefont {Biella},
  \citenamefont {Borgonovi}, \citenamefont {Kaiser},\ and\ \citenamefont
  {Celardo}}]{Biella2013}%
  \BibitemOpen
  \bibfield  {author} {\bibinfo {author} {\bibfnamefont {A.}~\bibnamefont
  {Biella}}, \bibinfo {author} {\bibfnamefont {F.}~\bibnamefont {Borgonovi}},
  \bibinfo {author} {\bibfnamefont {R.}~\bibnamefont {Kaiser}}, \ and\ \bibinfo
  {author} {\bibfnamefont {G.~L.}\ \bibnamefont {Celardo}},\ }\bibfield
  {title} {\enquote {\bibinfo {title} {Subradiant hybrid states in the open 3d
  anderson-dicke model},}\ }\href {\doibase 10.1209/0295-5075/103/57009}
  {\bibfield  {journal} {\bibinfo  {journal} {Europhys. Lett.}\ }\textbf
  {\bibinfo {volume} {103}},\ \bibinfo {pages} {57009} (\bibinfo {year}
  {2013})}\BibitemShut {NoStop}%
\bibitem [{Note1()}]{Note1}%
  \BibitemOpen
  \bibinfo {note} {Note that $\protect \mathaccentV {bar}016{V}_{\alpha \alpha
  }=0$ and $\protect \mathaccentV {bar}016{\Delta }_{\alpha \beta }=\protect
  \mathaccentV {bar}016{\Delta }_{\alpha \alpha }\delta _{\alpha \beta
  }$}\BibitemShut {NoStop}%
\bibitem [{Note2()}]{Note2}%
  \BibitemOpen
  \bibinfo {note} {Note that this approximation only holds for small enough
  values of $a/\lambda $. As this ratio is increased, the decay rate $\Gamma
  _k$ of the subradiant states get closer to $\gamma $ and hence the plateau
  only holds for a small period of time $\tau $, given by the difference
  between $1/\Gamma _k$ and $1/\Gamma _{k+1}$.}\BibitemShut {Stop}%
\bibitem [{\citenamefont {Yamamoto}\ \emph {et~al.}(2016)\citenamefont
  {Yamamoto}, \citenamefont {Kobayashi}, \citenamefont {Kuno}, \citenamefont
  {Kato},\ and\ \citenamefont {Takahashi}}]{Yamamoto2016}%
  \BibitemOpen
  \bibfield  {author} {\bibinfo {author} {\bibfnamefont {Ryuta}\ \bibnamefont
  {Yamamoto}}, \bibinfo {author} {\bibfnamefont {Jun}\ \bibnamefont
  {Kobayashi}}, \bibinfo {author} {\bibfnamefont {Takuma}\ \bibnamefont
  {Kuno}}, \bibinfo {author} {\bibfnamefont {Kohei}\ \bibnamefont {Kato}}, \
  and\ \bibinfo {author} {\bibfnamefont {Yoshiro}\ \bibnamefont {Takahashi}},\
  }\bibfield  {title} {\enquote {\bibinfo {title} {{An ytterbium quantum gas
  microscope with narrow-line laser cooling}},}\ }\href {\doibase
  10.1088/1367-2630/18/2/023016} {\bibfield  {journal} {\bibinfo  {journal}
  {New J. Phys.}\ }\textbf {\bibinfo {volume} {18}},\ \bibinfo {pages} {023016}
  (\bibinfo {year} {2016})},\ \Eprint {http://arxiv.org/abs/1509.03233}
  {arXiv:1509.03233} \BibitemShut {NoStop}%
\bibitem [{\citenamefont {Miranda}\ \emph {et~al.}(2017)\citenamefont
  {Miranda}, \citenamefont {Inoue}, \citenamefont {Tambo},\ and\ \citenamefont
  {Kozuma}}]{Miranda2017}%
  \BibitemOpen
  \bibfield  {author} {\bibinfo {author} {\bibfnamefont {Martin}\ \bibnamefont
  {Miranda}}, \bibinfo {author} {\bibfnamefont {Ryotaro}\ \bibnamefont
  {Inoue}}, \bibinfo {author} {\bibfnamefont {Naoki}\ \bibnamefont {Tambo}}, \
  and\ \bibinfo {author} {\bibfnamefont {Mikio}\ \bibnamefont {Kozuma}},\
  }\bibfield  {title} {\enquote {\bibinfo {title} {{Site-resolved imaging of a
  bosonic Mott insulator using ytterbium atoms}},}\ }\href {\doibase
  10.1103/PhysRevA.96.043626} {\bibfield  {journal} {\bibinfo  {journal} {Phys.
  Rev. A}\ }\textbf {\bibinfo {volume} {96}},\ \bibinfo {pages} {043626}
  (\bibinfo {year} {2017})},\ \Eprint {http://arxiv.org/abs/1704.07060}
  {arXiv:1704.07060} \BibitemShut {NoStop}%
\bibitem [{\citenamefont {Norcia}\ \emph {et~al.}(2018)\citenamefont {Norcia},
  \citenamefont {Young},\ and\ \citenamefont {Kaufman}}]{Norcia2018}%
  \BibitemOpen
  \bibfield  {author} {\bibinfo {author} {\bibfnamefont {M.~A.}\ \bibnamefont
  {Norcia}}, \bibinfo {author} {\bibfnamefont {A.~W.}\ \bibnamefont {Young}}, \
  and\ \bibinfo {author} {\bibfnamefont {A.~M.}\ \bibnamefont {Kaufman}},\
  }\bibfield  {title} {\enquote {\bibinfo {title} {Microscopic control and
  detection of ultracold strontium in optical-tweezer arrays},}\ }\href
  {\doibase 10.1103/PhysRevX.8.041054} {\bibfield  {journal} {\bibinfo
  {journal} {Phys. Rev. X}\ }\textbf {\bibinfo {volume} {8}},\ \bibinfo {pages}
  {041054} (\bibinfo {year} {2018})}\BibitemShut {NoStop}%
\bibitem [{\citenamefont {Jones}\ \emph {et~al.}(2018)\citenamefont {Jones},
  \citenamefont {Needham}, \citenamefont {Lesanovsky}, \citenamefont
  {Intravaia},\ and\ \citenamefont {Olmos}}]{Jones2018}%
  \BibitemOpen
  \bibfield  {author} {\bibinfo {author} {\bibfnamefont {Ryan}\ \bibnamefont
  {Jones}}, \bibinfo {author} {\bibfnamefont {Jemma~A.}\ \bibnamefont
  {Needham}}, \bibinfo {author} {\bibfnamefont {Igor}\ \bibnamefont
  {Lesanovsky}}, \bibinfo {author} {\bibfnamefont {Francesco}\ \bibnamefont
  {Intravaia}}, \ and\ \bibinfo {author} {\bibfnamefont {Beatriz}\ \bibnamefont
  {Olmos}},\ }\bibfield  {title} {\enquote {\bibinfo {title} {{Modified
  dipole-dipole interaction and dissipation in an atomic ensemble near
  surfaces}},}\ }\href {\doibase 10.1103/PhysRevA.97.053841} {\bibfield
  {journal} {\bibinfo  {journal} {Phys. Rev. A}\ }\textbf {\bibinfo {volume}
  {97}},\ \bibinfo {pages} {053841} (\bibinfo {year} {2018})},\ \Eprint
  {http://arxiv.org/abs/1803.06309} {arXiv:1803.06309} \BibitemShut {NoStop}%
\bibitem [{\citenamefont {Le~Kien}\ and\ \citenamefont
  {Rauschenbeutel}(2017)}]{LeKien2017}%
  \BibitemOpen
  \bibfield  {author} {\bibinfo {author} {\bibfnamefont {Fam}\ \bibnamefont
  {Le~Kien}}\ and\ \bibinfo {author} {\bibfnamefont {A.}~\bibnamefont
  {Rauschenbeutel}},\ }\bibfield  {title} {\enquote {\bibinfo {title}
  {Nanofiber-mediated chiral radiative coupling between two atoms},}\ }\href
  {\doibase 10.1103/PhysRevA.95.023838} {\bibfield  {journal} {\bibinfo
  {journal} {Phys. Rev. A}\ }\textbf {\bibinfo {volume} {95}},\ \bibinfo
  {pages} {023838} (\bibinfo {year} {2017})}\BibitemShut {NoStop}%
\bibitem [{\citenamefont {Lodahl}\ \emph {et~al.}(2017)\citenamefont {Lodahl},
  \citenamefont {Mahmoodian}, \citenamefont {Stobbe}, \citenamefont
  {Rauschenbeutel}, \citenamefont {Schneeweiss}, \citenamefont {Volz},
  \citenamefont {Pichler},\ and\ \citenamefont {Zoller}}]{Lodahl2017}%
  \BibitemOpen
  \bibfield  {author} {\bibinfo {author} {\bibfnamefont {Peter}\ \bibnamefont
  {Lodahl}}, \bibinfo {author} {\bibfnamefont {Sahand}\ \bibnamefont
  {Mahmoodian}}, \bibinfo {author} {\bibfnamefont {S{\o}ren}\ \bibnamefont
  {Stobbe}}, \bibinfo {author} {\bibfnamefont {Arno}\ \bibnamefont
  {Rauschenbeutel}}, \bibinfo {author} {\bibfnamefont {Philipp}\ \bibnamefont
  {Schneeweiss}}, \bibinfo {author} {\bibfnamefont {J{\"u}rgen}\ \bibnamefont
  {Volz}}, \bibinfo {author} {\bibfnamefont {Hannes}\ \bibnamefont {Pichler}},
  \ and\ \bibinfo {author} {\bibfnamefont {Peter}\ \bibnamefont {Zoller}},\
  }\bibfield  {title} {\enquote {\bibinfo {title} {Chiral quantum optics},}\
  }\href {https://doi.org/10.1038/nature21037} {\bibfield  {journal} {\bibinfo
  {journal} {Nature}\ }\textbf {\bibinfo {volume} {541}},\ \bibinfo {pages}
  {473} (\bibinfo {year} {2017})}\BibitemShut {NoStop}%
\bibitem [{\citenamefont {Buonaiuto}\ \emph {et~al.}(2019)\citenamefont
  {Buonaiuto}, \citenamefont {Jones}, \citenamefont {Olmos},\ and\
  \citenamefont {Lesanovsky}}]{Buonaiuto2019}%
  \BibitemOpen
  \bibfield  {author} {\bibinfo {author} {\bibfnamefont {Giuseppe}\
  \bibnamefont {Buonaiuto}}, \bibinfo {author} {\bibfnamefont {Ryan}\
  \bibnamefont {Jones}}, \bibinfo {author} {\bibfnamefont {Beatriz}\
  \bibnamefont {Olmos}}, \ and\ \bibinfo {author} {\bibfnamefont {Igor}\
  \bibnamefont {Lesanovsky}},\ }\bibfield  {title} {\enquote {\bibinfo {title}
  {Dynamical creation and detection of entangled many-body states in a chiral
  atom chain},}\ }\href@noop {} {\bibfield  {journal} {\bibinfo  {journal}
  {arXiv:1902.08525}\ } (\bibinfo {year} {2019})}\BibitemShut {NoStop}%
\bibitem [{\citenamefont {Albrecht}\ \emph {et~al.}(2019)\citenamefont
  {Albrecht}, \citenamefont {Henriet}, \citenamefont {Asenjo-Garcia},
  \citenamefont {Dieterle}, \citenamefont {Painter},\ and\ \citenamefont
  {Chang}}]{Albrecht2019}%
  \BibitemOpen
  \bibfield  {author} {\bibinfo {author} {\bibfnamefont {Andreas}\ \bibnamefont
  {Albrecht}}, \bibinfo {author} {\bibfnamefont {Loïc}\ \bibnamefont
  {Henriet}}, \bibinfo {author} {\bibfnamefont {Ana}\ \bibnamefont
  {Asenjo-Garcia}}, \bibinfo {author} {\bibfnamefont {Paul~B}\ \bibnamefont
  {Dieterle}}, \bibinfo {author} {\bibfnamefont {Oskar}\ \bibnamefont
  {Painter}}, \ and\ \bibinfo {author} {\bibfnamefont {Darrick~E}\ \bibnamefont
  {Chang}},\ }\bibfield  {title} {\enquote {\bibinfo {title} {Subradiant states
  of quantum bits coupled to a one-dimensional waveguide},}\ }\href {\doibase
  10.1088/1367-2630/ab0134} {\bibfield  {journal} {\bibinfo  {journal} {New J.
  Phys.}\ }\textbf {\bibinfo {volume} {21}},\ \bibinfo {pages} {025003}
  (\bibinfo {year} {2019})}\BibitemShut {NoStop}%
\bibitem [{\citenamefont {Weitenberg}\ \emph {et~al.}(2011)\citenamefont
  {Weitenberg}, \citenamefont {Endres}, \citenamefont {Sherson}, \citenamefont
  {Cheneau}, \citenamefont {Schauß}, \citenamefont {Fukuhara}, \citenamefont
  {Bloch},\ and\ \citenamefont {Kuhr}}]{Weitenberg2011}%
  \BibitemOpen
  \bibfield  {author} {\bibinfo {author} {\bibfnamefont {Christof}\
  \bibnamefont {Weitenberg}}, \bibinfo {author} {\bibfnamefont {Manuel}\
  \bibnamefont {Endres}}, \bibinfo {author} {\bibfnamefont {Jacob~F.}\
  \bibnamefont {Sherson}}, \bibinfo {author} {\bibfnamefont {Marc}\
  \bibnamefont {Cheneau}}, \bibinfo {author} {\bibfnamefont {Peter}\
  \bibnamefont {Schauß}}, \bibinfo {author} {\bibfnamefont {Takeshi}\
  \bibnamefont {Fukuhara}}, \bibinfo {author} {\bibfnamefont {Immanuel}\
  \bibnamefont {Bloch}}, \ and\ \bibinfo {author} {\bibfnamefont {Stefan}\
  \bibnamefont {Kuhr}},\ }\bibfield  {title} {\enquote {\bibinfo {title}
  {Single-spin addressing in an atomic mott insulator},}\ }\href
  {https://doi.org/10.1038/nature09827} {\bibfield  {journal} {\bibinfo
  {journal} {Nature}\ }\textbf {\bibinfo {volume} {471}},\ \bibinfo {pages}
  {319} (\bibinfo {year} {2011})}\BibitemShut {NoStop}%
\bibitem [{\citenamefont {Kim}\ \emph {et~al.}(2016)\citenamefont {Kim},
  \citenamefont {Lee}, \citenamefont {Jo}, \citenamefont {Song},\ and\
  \citenamefont {Ahn}}]{Kim2016}%
  \BibitemOpen
  \bibfield  {author} {\bibinfo {author} {\bibfnamefont {Hyosub}\ \bibnamefont
  {Kim}}, \bibinfo {author} {\bibfnamefont {Han-gyeol}\ \bibnamefont {Lee},
  \bibfnamefont {Woojun~andLee}}, \bibinfo {author} {\bibfnamefont {Hanlae}\
  \bibnamefont {Jo}}, \bibinfo {author} {\bibfnamefont {Yunheung}\ \bibnamefont
  {Song}}, \ and\ \bibinfo {author} {\bibfnamefont {Jaewook}\ \bibnamefont
  {Ahn}},\ }\bibfield  {title} {\enquote {\bibinfo {title} {In situ single-atom
  array synthesis using dynamic holographic optical tweezers},}\ }\href
  {https://doi.org/10.1038/ncomms13317} {\bibfield  {journal} {\bibinfo
  {journal} {Nat. Comm.}\ }\textbf {\bibinfo {volume} {7}},\ \bibinfo {pages}
  {13317} (\bibinfo {year} {2016})}\BibitemShut {NoStop}%
\bibitem [{\citenamefont {Endres}\ \emph {et~al.}(2016)\citenamefont {Endres},
  \citenamefont {Bernien}, \citenamefont {Keesling}, \citenamefont {Levine},
  \citenamefont {Anschuetz}, \citenamefont {Krajenbrink}, \citenamefont
  {Senko}, \citenamefont {Vuletic}, \citenamefont {Greiner},\ and\
  \citenamefont {Lukin}}]{Endres2016}%
  \BibitemOpen
  \bibfield  {author} {\bibinfo {author} {\bibfnamefont {Manuel}\ \bibnamefont
  {Endres}}, \bibinfo {author} {\bibfnamefont {Hannes}\ \bibnamefont
  {Bernien}}, \bibinfo {author} {\bibfnamefont {Alexander}\ \bibnamefont
  {Keesling}}, \bibinfo {author} {\bibfnamefont {Harry}\ \bibnamefont
  {Levine}}, \bibinfo {author} {\bibfnamefont {Eric~R.}\ \bibnamefont
  {Anschuetz}}, \bibinfo {author} {\bibfnamefont {Alexandre}\ \bibnamefont
  {Krajenbrink}}, \bibinfo {author} {\bibfnamefont {Crystal}\ \bibnamefont
  {Senko}}, \bibinfo {author} {\bibfnamefont {Vladan}\ \bibnamefont {Vuletic}},
  \bibinfo {author} {\bibfnamefont {Markus}\ \bibnamefont {Greiner}}, \ and\
  \bibinfo {author} {\bibfnamefont {Mikhail~D.}\ \bibnamefont {Lukin}},\
  }\bibfield  {title} {\enquote {\bibinfo {title} {Atom-by-atom assembly of
  defect-free one-dimensional cold atom arrays},}\ }\href {\doibase
  10.1126/science.aah3752} {\bibfield  {journal} {\bibinfo  {journal}
  {Science}\ }\textbf {\bibinfo {volume} {354}},\ \bibinfo {pages} {1024}
  (\bibinfo {year} {2016})}\BibitemShut {NoStop}%
\bibitem [{\citenamefont {Barredo}\ \emph {et~al.}(2016)\citenamefont
  {Barredo}, \citenamefont {de~L{\'e}s{\'e}leuc}, \citenamefont {Lienhard},
  \citenamefont {Lahaye},\ and\ \citenamefont {Browaeys}}]{Barredo2016}%
  \BibitemOpen
  \bibfield  {author} {\bibinfo {author} {\bibfnamefont {Daniel}\ \bibnamefont
  {Barredo}}, \bibinfo {author} {\bibfnamefont {Sylvain}\ \bibnamefont
  {de~L{\'e}s{\'e}leuc}}, \bibinfo {author} {\bibfnamefont {Vincent}\
  \bibnamefont {Lienhard}}, \bibinfo {author} {\bibfnamefont {Thierry}\
  \bibnamefont {Lahaye}}, \ and\ \bibinfo {author} {\bibfnamefont {Antoine}\
  \bibnamefont {Browaeys}},\ }\bibfield  {title} {\enquote {\bibinfo {title}
  {An atom-by-atom assembler of defect-free arbitrary two-dimensional atomic
  arrays},}\ }\href {\doibase 10.1126/science.aah3778} {\bibfield  {journal}
  {\bibinfo  {journal} {Science}\ }\textbf {\bibinfo {volume} {354}},\ \bibinfo
  {pages} {1021} (\bibinfo {year} {2016})},\ \Eprint
  {http://arxiv.org/abs/http://science.sciencemag.org/content/354/6315/1021.full.pdf}
  {http://science.sciencemag.org/content/354/6315/1021.full.pdf} \BibitemShut
  {NoStop}%
\bibitem [{\citenamefont {Cooper}\ \emph {et~al.}(2018)\citenamefont {Cooper},
  \citenamefont {Covey}, \citenamefont {Madjarov}, \citenamefont {Porsev},
  \citenamefont {Safronova},\ and\ \citenamefont {Endres}}]{Cooper2018}%
  \BibitemOpen
  \bibfield  {author} {\bibinfo {author} {\bibfnamefont {Alexandre}\
  \bibnamefont {Cooper}}, \bibinfo {author} {\bibfnamefont {Jacob~P.}\
  \bibnamefont {Covey}}, \bibinfo {author} {\bibfnamefont {Ivaylo~S.}\
  \bibnamefont {Madjarov}}, \bibinfo {author} {\bibfnamefont {Sergey~G.}\
  \bibnamefont {Porsev}}, \bibinfo {author} {\bibfnamefont {Marianna~S.}\
  \bibnamefont {Safronova}}, \ and\ \bibinfo {author} {\bibfnamefont {Manuel}\
  \bibnamefont {Endres}},\ }\bibfield  {title} {\enquote {\bibinfo {title}
  {Alkaline-earth atoms in optical tweezers},}\ }\href {\doibase
  10.1103/PhysRevX.8.041055} {\bibfield  {journal} {\bibinfo  {journal} {Phys.
  Rev. X}\ }\textbf {\bibinfo {volume} {8}},\ \bibinfo {pages} {041055}
  (\bibinfo {year} {2018})}\BibitemShut {NoStop}%
\bibitem [{\citenamefont {Zhang}\ and\ \citenamefont
  {M{\o}lmer}(2019)}]{Zhang2019}%
  \BibitemOpen
  \bibfield  {author} {\bibinfo {author} {\bibfnamefont {Yu-Xiang}\
  \bibnamefont {Zhang}}\ and\ \bibinfo {author} {\bibfnamefont {Klaus}\
  \bibnamefont {M{\o}lmer}},\ }\bibfield  {title} {\enquote {\bibinfo {title}
  {Theory of subradiant states of a one-dimensional two-level atom chain},}\
  }\href@noop {} {\bibfield  {journal} {\bibinfo  {journal} {arXiv:1812.09784}\
  } (\bibinfo {year} {2019})}\BibitemShut {NoStop}%
\bibitem [{\citenamefont {Gorshkov}\ \emph {et~al.}(2011)\citenamefont
  {Gorshkov}, \citenamefont {Otterbach}, \citenamefont {Fleischhauer},
  \citenamefont {Pohl},\ and\ \citenamefont {Lukin}}]{Gorshkov2011}%
  \BibitemOpen
  \bibfield  {author} {\bibinfo {author} {\bibfnamefont {Alexey~V.}\
  \bibnamefont {Gorshkov}}, \bibinfo {author} {\bibfnamefont {Johannes}\
  \bibnamefont {Otterbach}}, \bibinfo {author} {\bibfnamefont {Michael}\
  \bibnamefont {Fleischhauer}}, \bibinfo {author} {\bibfnamefont {Thomas}\
  \bibnamefont {Pohl}}, \ and\ \bibinfo {author} {\bibfnamefont {Mikhail~D.}\
  \bibnamefont {Lukin}},\ }\bibfield  {title} {\enquote {\bibinfo {title}
  {Photon-photon interactions via rydberg blockade},}\ }\href {\doibase
  10.1103/PhysRevLett.107.133602} {\bibfield  {journal} {\bibinfo  {journal}
  {Phys. Rev. Lett.}\ }\textbf {\bibinfo {volume} {107}},\ \bibinfo {pages}
  {133602} (\bibinfo {year} {2011})}\BibitemShut {NoStop}%
\bibitem [{\citenamefont {Tiarks}\ \emph {et~al.}(2019)\citenamefont {Tiarks},
  \citenamefont {Schmidt-Eberle}, \citenamefont {Stolz}, \citenamefont
  {Rempe},\ and\ \citenamefont {D\"urr}}]{Tiarks2019}%
  \BibitemOpen
  \bibfield  {author} {\bibinfo {author} {\bibfnamefont {Daniel}\ \bibnamefont
  {Tiarks}}, \bibinfo {author} {\bibfnamefont {Steffen}\ \bibnamefont
  {Schmidt-Eberle}}, \bibinfo {author} {\bibfnamefont {Thomas}\ \bibnamefont
  {Stolz}}, \bibinfo {author} {\bibfnamefont {Gerhard}\ \bibnamefont {Rempe}},
  \ and\ \bibinfo {author} {\bibfnamefont {Stephan}\ \bibnamefont {D\"urr}},\
  }\bibfield  {title} {\enquote {\bibinfo {title} {A photon–photon quantum
  gate based on rydberg interactions},}\ }\href@noop {} {\bibfield  {journal}
  {\bibinfo  {journal} {Nat. Phys.}\ }\textbf {\bibinfo {volume} {15}},\
  \bibinfo {pages} {124} (\bibinfo {year} {2019})}\BibitemShut {NoStop}%
\end{thebibliography}
\end{document}